\documentclass{cta-author}

\usepackage{graphicx,bm,natbib}
\usepackage{amsbsy,amsmath,mathrsfs}
\usepackage[dvipsnames]{xcolor}
\usepackage[T1]{fontenc}
\usepackage{newtxtext,newtxmath}

\usepackage[scaled]{helvet}
\usepackage{tikz}
\usepackage[normalem]{ulem}
\usepackage[caption=false]{subfig}
\usepackage{tabularx,makecell,xfrac}

\usetikzlibrary{arrows,decorations.pathmorphing,backgrounds,positioning,fit,petri}
\usepackage{gensymb}

\usepackage{amsthm}

\DeclareGraphicsExtensions{.pdf, .jpg, .eps, .svg}

\definecolor{RoyalAzure}{rgb}{0.0, 0.22, 0.66}
\usepackage[colorlinks=true,citecolor=RoyalAzure,linkcolor=RoyalAzure,urlcolor=RoyalAzure]{hyperref}

\definecolor{customcol}{rgb}{.1,0.7,0.1}

\begin{document}

\supertitle{This paper is a preprint of a paper submitted to IET Quantum Communication. If accepted, the copy of record will be available at the IET Digital Library}

\title{Advances in Space Quantum Communications}

\author{
\au{Jasminder S. Sidhu$^{1\corr}$},
\au{Siddarth K. Joshi$^{2}$},
\au{Mustafa G\"{u}ndo\u{g}an$^{3}$},
\au{Thomas Brougham$^{1}$},
\au{David Lowndes$^{2}$},
\au{Luca Mazzarella$^{4}$},
\au{Markus Krutzik$^{3}$}, 
\au{Sonali Mohapatra$^{1,14}$}, 
\au{Daniele Dequal$^{5}$},
\au{Giuseppe Vallone$^{6,7}$},
\au{Paolo Villoresi$^{6}$},
\au{Alexander Ling$^{8,9}$},
\au{Thomas Jennewein$^{10,11}$}
\au{Makan Mohageg$^{4}$},
\au{John Rarity$^{2}$},
\au{Ivette Fuentes$^{12}$},
\au{Stefano Pirandola$^{13}$},
\au{Daniel K. L. Oi$^{1}$}
}

\address{
\add{1}{SUPA Department of Physics, University of Strathclyde, 107 Rottenrow, Glasgow G4 0NG, United Kingdom}
\add{2}{Quantum Engineering Technology Labs, H. H. Wills Physics Laboratory \& Department of Electrical and
Electronic Engineering, University of Bristol, Merchant Venturers Building, Woodland Road, Bristol BS8
1UB, United Kingdom}
\add{3}{Instit\"{u}t f\"{u}r Physik, Humboldt-Universit\"{a}t zu Berlin, Newtonstr. 15, Berlin 12489, Germany}
\add{4}{Jet Propulsion Laboratory, California Institute of Technology, USA}
\add{5}{Scientific Research Unit, Agenzia Spaziale Italiana, Matera, Italy}
\add{6}{Dipartimento di Ingegneria dell'Informazione, Universit\`{a} degli Studi di Padova, Padova 35131, Italy}
\add{7}{Dipartimento di Fisica e Astronomia, Universit\`{a} degli Studi di Padova, Padova 35131, Italy}
\add{8}{Centre for Quantum Technologies, National University of Singapore, S15-02-10, 3 Science Drive 2, Singapore 117543, Singapore}
\add{9}{Department of Physics, 2 Science Drive 3 Blk S12, Level 2, Singapore 117551, Singapore}
\add{10}{Institute for Quantum Computing, University of Waterloo, Waterloo, ON, N2L 3G1 Canada}
\add{11}{Department of Physics and Astronomy, University of Waterloo, Waterloo, ON, N2L 3G1 Canada}
\add{12}{School of Physics and Astronomy, University of Southampton, Southampton SO17 1BJ, United Kingdom}
\add{13}{Department of Computer Science, University of York, York YO10 5GH, United Kingdom}
\add{14}{Craft Prospect Ltd., Suite 4A, Fairfield
1048, Govan Road, Glasgow, G51 4XS, United Kingdom}
\email{jasminder.sidhu@strath.ac.uk}
}

\begin{abstract}
Concerted efforts are underway to establish an infrastructure for a global quantum internet to realise a spectrum of quantum technologies. This will enable more precise sensors, secure communications, and faster data processing. Quantum communications are a front-runner with quantum networks already implemented in several metropolitan areas. A number of recent proposals have modelled the use of space segments to overcome range limitations of purely terrestrial networks. Rapid progress in the design of quantum devices have enabled their deployment in space for in-orbit demonstrations. We review developments in this emerging area of space-based quantum technologies and provide a roadmap of key milestones towards a complete, global quantum networked landscape. Small satellites hold increasing promise to provide a cost effective coverage required to realised the quantum internet. We review the state of art in small satellite missions and collate the most current in-field demonstrations of quantum cryptography. We summarise important challenges in space quantum technologies that must be overcome and recent efforts to mitigate their effects. A perspective on future developments that would improve the performance of space quantum communications is included. We conclude with a discussion on fundamental physics experiments that could take advantage of a global, space-based quantum network.
\end{abstract}

\maketitle

\section{Introduction}
\label{sec1}

\noindent
The second quantum revolution promises a paradigm shift in quantum technologies that will deliver new capabilities, and performance enhancements in security, accuracy, and precision~\cite{dowling2003quantum}. The working principles that permit these performance improvements are deep-rooted phenomena in quantum mechanics, such as entanglement~\cite{Horodecki2009_RMP}, teleportation~\cite{Bennett1993_PRL, Braunstein1998_PRL, Pirandola2015_NP}, the Heisenberg uncertainty principle~\cite{Heisenberg1927} and the no-cloning theorem~\cite{Wootters1982_N}. With no classical analogues, these effects illustrate a key departure from classical physics and are at the heart of delivering improvements to technologies that underpin modern society. Fundamentally, quantum theory governs how information is instantiated and processed in the configuration of matter and energy. This realisation led to an explosion in the development of quantum technologies. The race to develop mature quantum technologies has become a discipline in its own right, with major renewed international efforts, both governmental and commercial.

A number of technologies can inherit enhanced performances and securities when availed to quantum mechanical resources. These technologies include sensing~\cite{Sidhu2017_PRA, Sidhu2018_arxiv,Rubio2020_JPA}, metrology~\cite{Sidhu2021_PRX, Sidhu2020_AVS,moreau2017demonstrating}, navigation and timing~\cite{Jozsa2000_PRL}, state discrimination~\cite{Helstrom1976,Peres1998_JPA, Sidhu2020_PRXQ}, communication~\cite{QKDreview2020,Scarani2009_RMP}, and computation~\cite{Knill2001_N}. Today, these technologies have long evolved from theoretical curiosities to significant developments and even field realisations. Notably, enhanced quantum sensors are now in the vanguard of early practical application of quantum technologies. Fundamental precision bounds to measurements of physical signals can be attained by exploiting specific quantum correlations in probe states~\cite{Braun2018_RMP}. This was recently applied in the Laser Interferometer Gravitational-Wave Observatory, which significantly improves the maturity of gravitational wave astronomy~\cite{Schnabel2010_NC, Ligo2013_NP, Dobrzanski2013_PRA}. Further, implementation of secure quantum communications have gained momentum owing to sustained improvements in quantum key distribution (QKD) protocols~\cite{QKDreview2020}. These improvements have boosted the appetite for early adoption of dedicated quantum technologies.

While focused works have demonstrated performance improvements across the spectrum of quantum technologies, they do not yet satisfy the concomitant technological demand. In addition, most practical applications require the joint operation of multiple quantum technologies. Recent works have established a path towards practical realisation and readiness, by requiring a networked infrastructure of dedicated quantum technologies to upscale capabilities. The essential resource in networked quantum technologies is quantum entanglement, which can be distributed directly~\cite{joshi2020trusted,wengerowsky2018entanglement, wengerowsky2020passively,wengerowsky2019entanglement} or through swapping~\cite{Zukowski1993_PRL, Bennett1993_PRL} and purification~\cite{Bennett1996_PRL,Deutsch1996_PRL} operations across an arbitrary topology of quantum nodes~\cite{Network_Capacities,Pirandola2019_CP,Pant2019_NPJQI,Network_Qzhuang,Das2018_PRA,Pirandola_2019Multi,Azuma2016_NC,Azuma2017_PRA,Network_Qzhuang2}. Heuristically, two attributes of quantum entanglement motivate a networked approach to quantum technologies. First, quantum correlations between entangled nodes grants maximum coordination among distant processors. This provides a natural approach to quantum clock synchronisation~\cite{Okeke2018_NPJQI,Jozsa2000_PRL}. It also heralds great potential for applications that employ distributed systems, such as distributed sensing~\cite{Zhuang2018_PRA,Humphreys2013_PRL, Komar2014_N, Eldredge2018_PRA, Polino2019_O,  Guo2020_NP}, astronomical long-baseline interferometry~\cite{Gottesman2012_PRL}, enhanced positioning~\cite{Qian2020_arxiv}, and consensus and agreement tasks~\cite{quantum_consensus,ByzantineAgreement2005}. Second, the monogamy of entanglement provides an inherent privacy that cannot be breached~\cite{Scarani2009_RMP}. This is particularly suited to quantum cryptography, where QKD enables secure communications between multiple remote network nodes~\cite{joshi2020trusted}. The impetus to develop quantum communication is primitive and born from the growing threat of quantum computers on the classical cryptosystems.

A networked infrastructure is also intuitively reasoned in quantum computing, where computationally intensive tasks are delegated across shared systems~\cite{Cirac1999_PRA,Fitzsimons2017_NPJQI}. In direct analogy with classical computing, the capability of quantum computers significantly improves with remote access to ubiquitous quantum processors. Improvements to this delegated approach to quantum computing is conditioned on the existence of high-speed global quantum communications networks. Viewed in this way, QKD has become the precursor to early applications of quantum communications. The parallel advancements in quantum communications and networked quantum computing progressively evolves the maturity of an emerging quantum internet~\cite{Kimble2008_N,Hybrid_qInternet,Wehner2018_S}. This is anticipated to impart a similar revolutionary impact on our world as the classical internet.

To date, the distance-limited span of quantum networks restricts a concrete implementation of this vision. Overcoming this limitation captures the essence of an increasing number of recent works in quantum communications. Ultimately, photonic losses dissolve the hope for a quantum internet based on optical fibre networks alone. Quantum repeaters~\cite{Briegel1998_PRL,Sangouard2011,Jiang2009_PRA,Munro2010_NP,Zwerger2018_PRL} can extend this capability by providing efficient routing of entanglement~\cite{Network_Capacities,Pirandola2019_CP, Pant2019_NPJQI,Network_Qzhuang}. Quantum networks have emerged in and between major metropolis areas~\cite{Castelvecchi2018_N, Dynes2019_NPJQI, joshi2020trusted, zhang2019quantum}. Despite significant progress on quantum repeaters, global quantum networks remain a hopeful endeavour. The inability to place quantum repeaters along geographically inimical locations, rules out option alone as insufficient. The emergence of the quantum internet necessarily requires the combination of both terrestrial and space-based segments.

In this review, we focus on capturing the fast progress in the emerging area of space-based quantum communications. We detail a roadmap to a future implementation of the quantum internet that unifies both terrestrial and space networks. We start by first exploring applications of a fully operational quantum network in section~\ref{sec:applications}. The state-of-art for each application is provided. In section~\ref{sec:current_efforts}, we review the progress to date in the emerging area of satellite-based quantum communications. This section highlights key milestones in field-demonstrations. We distinguish these demonstrations in terms of the size of the satellite(s) involved in the quantum network. In section~\ref{sec2:challenges}, we explore outstanding challenges that currently impede the maturity of space quantum communications. We draw attention to current efforts to mitigate these challenges. In section~\ref{sec:enabling_tech}, we review key enabling technologies. Section~\ref{sec:fundamental_expts} provides an outlook to the open questions in quantum mechanics that can be addressed with access to a mature realisation of space-based quantum communication network. Throughout this review, we aim to capture the rapid progress in the race to realise global quantum communications. To achieve this, we draw attention to progress in different academic, governmental, and commercial efforts, in addition to ongoing difficulties in space technologies.


\section{Applications}
\label{sec:applications}

\noindent
The quantum internet is set to deliver a profound impact across a range of technological frontiers, including quantum communications, computation, and metrology. The working infrastructure to this is a globally interconnected network of quantum information processors, that deliver enhanced capabilities over the use of purely classical information. Attaining such a connectivity over global scale requires the maturity of both ground and satellite nodes, that allows efficient distribution and routing of quantum entanglement across the network, and teleportation of quantum states between nodes.

\begin{figure}[t!]
\centering
\includegraphics[width =0.99\columnwidth]{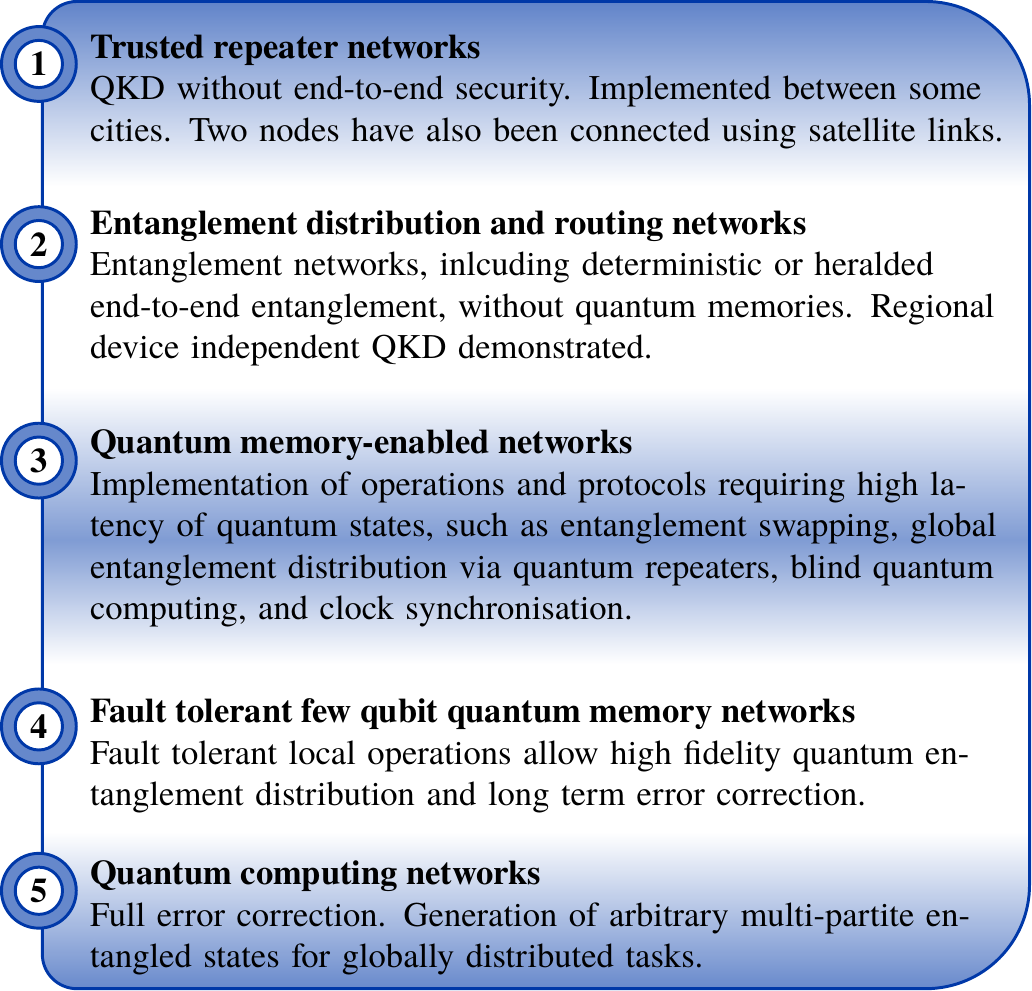}
\caption{Stages to a global satellite-based quantum internet. Demonstrating each state reflects an increase to the functionality of the network at the expense of increased technological difficulty. For each stage, we note example applications and protocols that can be demonstrated. See Ref.~\cite{Wehner2018_S} for a terrestrial-focused roadmap.}
\label{fig:qinternet_roadmap}
\end{figure}

A roadmap for the development of satellite-based quantum networks is presented in Fig.~\ref{fig:qinternet_roadmap}. In this section, we review specific applications of the quantum internet.


\subsection{Quantum key distribution}
\label{subsec:qkd}

\noindent
Quantum computation~\cite{Montanaro2016_NPJQI} has the potential for a paradigm change for solving problems in simulation and optimisation~\cite{McCaskey2019_NPJQI, Bauer2020_arxiv}. However, these new computational capabilities also threaten to the security of near-ubiquitous public key cryptosystems such as the Rivest-Shamir-Adleman (RSA) protocol~\cite{Rivest1978_CACM} or Elliptic Curve Cryptography~\cite{ecc_misc} that underlie much of internet infrastructure. While purely classical advances in number theory have posed ongoing threats to RSA~\cite{Agrawal2004_AM}, the possibility of implementing Shor’s algorithm~\cite{Shor1997_SIAM} on a large scale quantum computer in the near term poses a risk to both RSA and ECC that cannot be ignored, with major advances in scaling up quantum processors by companies such as Google~\cite{Boixo2018_NP}, IBM~\cite{Chamberland2020_PRX}, and Honeywell~\cite{Pino2020_arxiv} highlighting a shrinking horizon until the cryptographic quantum apocalypse. Alternative methods for private communication that are quantum secure need to be developed.


Progress along this endeavour has primarily proceeded along two distinct routes. First is post-quantum cryptography (PQC), which replaces RSA with alternative classical cryptosystems that are robust to factorisation and quantum algorithms~\cite{Bernstein2009_book}. However, PQC schemes provide a partial solution since they are not information-theoretically secure. A second more promising candidate for quantum-safe encryption is quantum key distribution (QKD), which guarantees the privacy, authentication, and confidentiality of secure communications. Specifically, signals encoded in optical quantum systems are processed to provide a secure encryption key, which comes with a computable assessment of the knowledge of an eavesdropper. QKD systems are operationally different to classical encryption systems. They take advantage of fundamental properties in quantum mechanics, such as the uncertainty principle or the monogamy of entanglement~\cite{Gisin2002_RMP,Scarani2009_RMP} to safeguard against adversaries with access to arbitrary computational power. There are three distinct families of different QKD protocols~\cite{QKDreview2020}. First are discrete-variable (DV) protocols that use discrete quantum degree of freedom to encode information, such as polarisation for free-space applications and phase coding in fiber-based approaches~\cite{Bennett1984_original, Bennett1992_PRL, Ekert1991_PRL, Boucher2005_PRA}. 

A second, natural alternative approach to QKD protocols is to use continuous-variable (CV) systems~\cite{CV-RMP,Ralph1999_PRA,Hillery2000_PRA,Reid2000_PRA,Gottesman2001_PRA, Cerf2001_PRA,Weedbrook2010}. CV systems can transfer more information per signal compared with qubit-based approaches and rely on cheaper technological implementations. A final class of QKD protocols, referred to as distributed-phase-reference coding,  implements a hybrid approach~\cite{Inoue2002_PRL, Inoue2003_PRA}. These protocols differ in the detection scheme used. DV and distributed-phase-reference protocols use photon counters and postselect the events in which a detection has effectively taken place. For some wavelengths, photon counters can suffer from low quantum efficiencies and are susceptible to high dark count rates and long dead times. CV protocols overcome these by using homodyne detection. However a tradeoff between key rate and noise must be addressed for CV approaches. QKD devices have continuously increased their key generation rate and have started approaching maturity, ready for implementation in realistic settings~\cite{Scarani2009_RMP, QKDreview2020}. Secure communications then follow through applications of the one-time pad encryption~\cite{Vernam1926, Miller1882_book}. This encryption scheme is provably secure provided the keys are secure and not reused, and they are as long as the text to be encrypted~\cite{Shannon1949_BSTJ}.

Prominent QKD protocols include the original Bennett-Brassard 1984 (BB84) protocol~\cite{Bennett1984_original}, the two-state Bennett 1992, (B92) protocol~\cite{Bennett92} and the six-state protocol~\cite{bruss1998}. For entanglement-based protocols, the common QKD protocols are the Ekert 91 (E91)~\cite{Ekert1991_PRL} and the BBM92~\cite{Bennett1992_PRL} protocols. Each protocol has their unique advantages and limitations. We refer the reader to Ref.~\cite{QKDreview2020} for a comprehensive review on different DV and CV protocols.

\begin{figure*}[t!]
\centering
\subfloat[Uplink channel.]{\includegraphics[width=0.3\columnwidth]{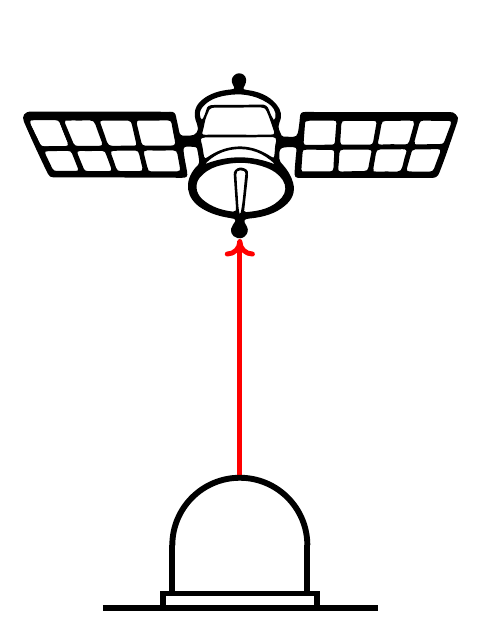}\label{fig:satlink1}} \hspace{15pt}
\subfloat[Downlink channel.]{\includegraphics[width=0.3\columnwidth]{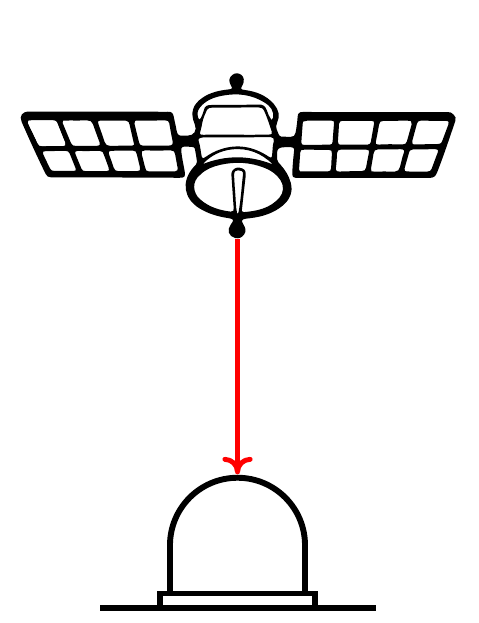} \label{fig:satlink2}} \hspace{15pt}
\subfloat[Double downlink.]{\includegraphics[width=0.358\columnwidth]{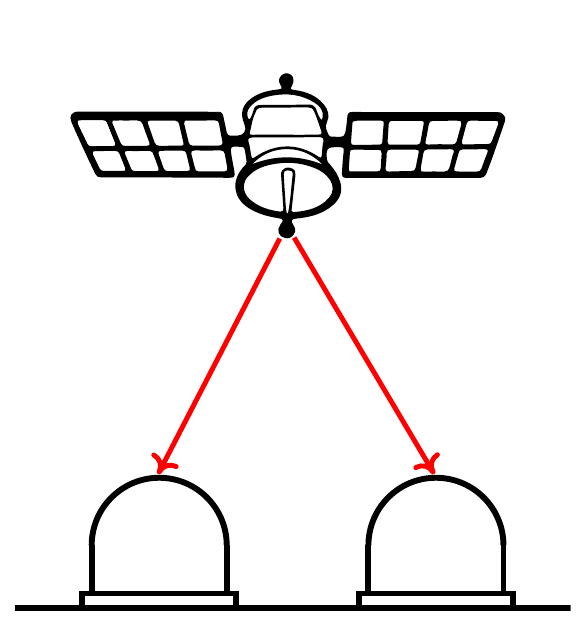} \label{fig:satlink3}} \hspace{15pt}
\subfloat[General inter-satellite/OGS links.]{\includegraphics[width=0.6\columnwidth]{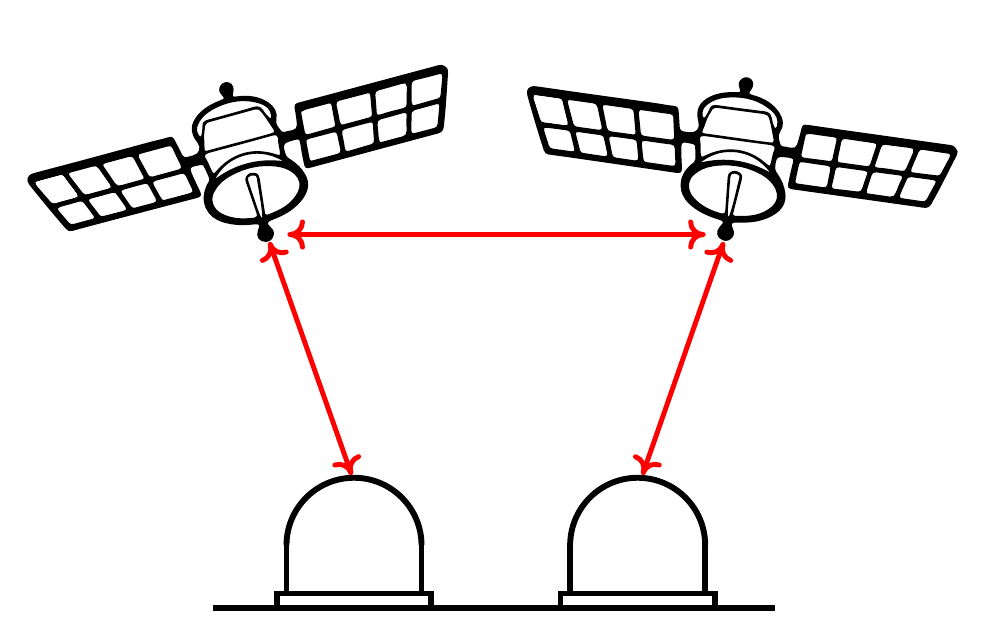} \label{fig:satlink4}}
\caption{Some satellite-based quantum communication configurations. There are many possible different ways of utilising satellites to mediate quantum communication links. Each red line represents a quantum channel from a transmitter ($T_x$) to a receiver ($R_x$). a) An optical ground station transmits quantum signal states to an orbiting satellite where they are measured~\cite{CSA_QEYSSat}. b) This topology can be reversed such that the satellite carries the quantum source and sends signals down to the Earth~\cite{Oi2017_EPJ}. c) A satellite distributes entanglement to two grounds stations by transmitting entangled photon pairs~\cite{Yin2017_S,Yin2020_N}. For MDI-QKD, a double uplink configuration could be used~\cite{tang2014experimental,cao2020long}. d) Inter-satellite links (ISLs) with ground links. In general, more complex networks topologies can be created from multiple channels~\cite{Gundogan2020_arxiv}.}
\label{fig:satellite_gs_links}
\end{figure*}%

The choice for the quantum degree of freedom used to encode information depends on the type of protocols used. A common DV approach is to use the polarisation of light~\cite{Bennett1984_original}. Polarisation states are simple to prepare, control and measure.  Furthermore, it is possible to produce high fidelity polarisation entangled photons~\cite{kwiat95}. Since there are only two linearly independent polarisation states, only one bit per photon can be encoded using this approach. There are alternative DV encodings that include the orbital angular momentum (OAM)~\cite{Mair2001,Mirhosseinil2015} and time-bin encodings~\cite{Alikhan2007,Islam2017}.  Both these approaches allow multiple encoded bits on each photon.  These high-dimensional encodings can also be more robust to certain types of errors~\cite{Nikolopoulos2006,Cerf2002}. However, these encodings are more difficult to realise experimentally. To illustrate this, consider the example of time-bin encodings. Entangled photon pairs that are in a superposition of many different discrete time-bins can be prepared using a mode-locked laser with beam-shaping techniques. It is possible to encode 10 bits per photon with practical laboratory conditions~\cite{Brougham2016}. However, the greater the number of time-bins, the more difficult it becomes to perform the mutually unbiased measurement needed to detect an eavesdropper.  For this reason, it is common to adopt a hybrid CV-DV encoding and use so-called energy-time entangled photons.  These  are photons that are entangled both in time-of-arrival and frequency.  Discrete time-bins can be imposed on the continuous time-of-arrival.  An eavesdropper is then be detected by a suitable interferometeric measurement~\cite{Zhong2015}.

QKD has been demonstrated reliably on the ground, and the process of being deployed across ground networks is well under way. To attain better key rates, practical implementations of QKD differ from theoretical proposals primarily in the components used in the system. Unfortunately, most security proofs are very sensitive to small differences between the physical devices used by the protocol and the theoretical model used to describe them. This introduces side channels to QKD systems that can be exploited for vulnerabilities in practical devices~\cite{Xu2020_RMP, Gisin2006_PRA, Zhao2008_PRA, Lydersen2010_NP, Xu2010_NJP, Gerhardt2011_NC, Weier2011_NJP, Huang2018_PRA_qkd}, therefore introducing an important trade-off between security and rate~\cite{Diamanti2016_NPJQI}. There is another crucial trade-off that must be considered. The foremost that hinders the maturity of QKD across global scales is the trade-off between rate and distance~\cite{ReverseCap,TakeokaGuha,PLOB}, typically induced by scattering effects in optical fibres~\cite{Brassard2000_PRL,Subacius2005_APL}. Specifically, the maximum number of secret bits that can be distributed over a lossy channel with transmissivity $\eta$ is upper bounded by the repeaterless PLOB bound $\smash{-\log_2(1-\eta)}$~\cite{PLOB}, which approximately amounts to $1.44 \eta$ bits per channel use at low transmissivity. This corresponds to the secret-key capacity of the lossy channel, an optimal key rate that cannot be overcome by any point-to-point protocol~\cite{QKDreview2020}. A similar limitation affects the free-space fading channel between two ground station, so that the optimal key rate cannot exceed the bound~\cite{FreeSpaceBounds} $\smash{-\Delta \log_2(1-\eta)}$, where $\eta$ is the maximum free-space line-of-sight transmissivity and $\Delta$ accounts for the non-trivial turbulence effects on the ground.

Quantum repeaters~\cite{Briegel1998_PRL,Sangouard2011,Jiang2009_PRA,Munro2010_NP,Zwerger2018_PRL} and, more generally, multi-hop quantum networks~\cite{Network_Capacities,Pirandola2019_CP, Pant2019_NPJQI,Network_Qzhuang} can provide better rates over long distances. However, these strategies provide only limited remission and have their own limitations in achievable key rates. In particular, for a chain of quantum repeaters connected by fibre-links with transmissivities $\eta_i$, the maximum secret key rate achievable by the two end-users (i.e., Alice and Bob at the two opposite ends of the chain) cannot exceed $-\log_2(1-\mathrm{min}_i \eta_i)$ bits per use of the chain~\cite{Pirandola2019_CP}. The previous formula implies that, for Alice and Bob being able to generate $1$ secret bit per chain use, we may tolerate at most 3 dB of loss in each individual link, which means that we need to insert a repeater every 15 km (at the standard fibre loss of $0.2$ dB/km). Therefore, despite significant progress, all ground-based QKD approaches remain distance-limited due to loss. These limitations make purely ground-based systems very challenging or even impractical for a global distribution network. Besides QKD also entanglement distribution remains an open challenge over global transmission lengths. This latter aspect, together with practical types of quantum repeaters, will be discussed in Sec.~\ref{subsec:entanglement_networks}.

The range of communication may be extended by employing satellites equipped with high-quality optical links. Fig.~\ref{fig:satellite_gs_links} illustrates different configurations of satellite-based quantum communications. The use of satellites reduces the demand on the number of ground quantum repeaters, since a single trusted-node satellite may physically travel from the proximity of Alice to the proximity of Bob, with the loss affecting only the uplink and downlink channels. It is also due to the fact that inter-satellite communications would be affected by less noise and loss with respect to ground links. The use of both ground and satellite-based quantum repeater networks provide the most promising solution to extend quantum communications to global scales. This provides a practical roadmap towards the implementation of a full quantum internet~\cite{Kimble2008_N,Hybrid_qInternet,Wehner2018_S}. It also paves the way to a promising realisation of a global networked infrastructure for global communication~\cite{Gundogan2020_arxiv, Liorni2020_arxiv}, imaging~\cite{Ciampini2016_SR}, and enhanced sensing~\cite{Sidhu2017_PRA, Sidhu2018_arxiv, Sidhu2021_PRX, Sidhu2020_AVS,Rubio2020_JPA}.

\begin{figure}[t]
\vspace{-0.0cm}
\centering
\includegraphics[width =0.99\columnwidth]{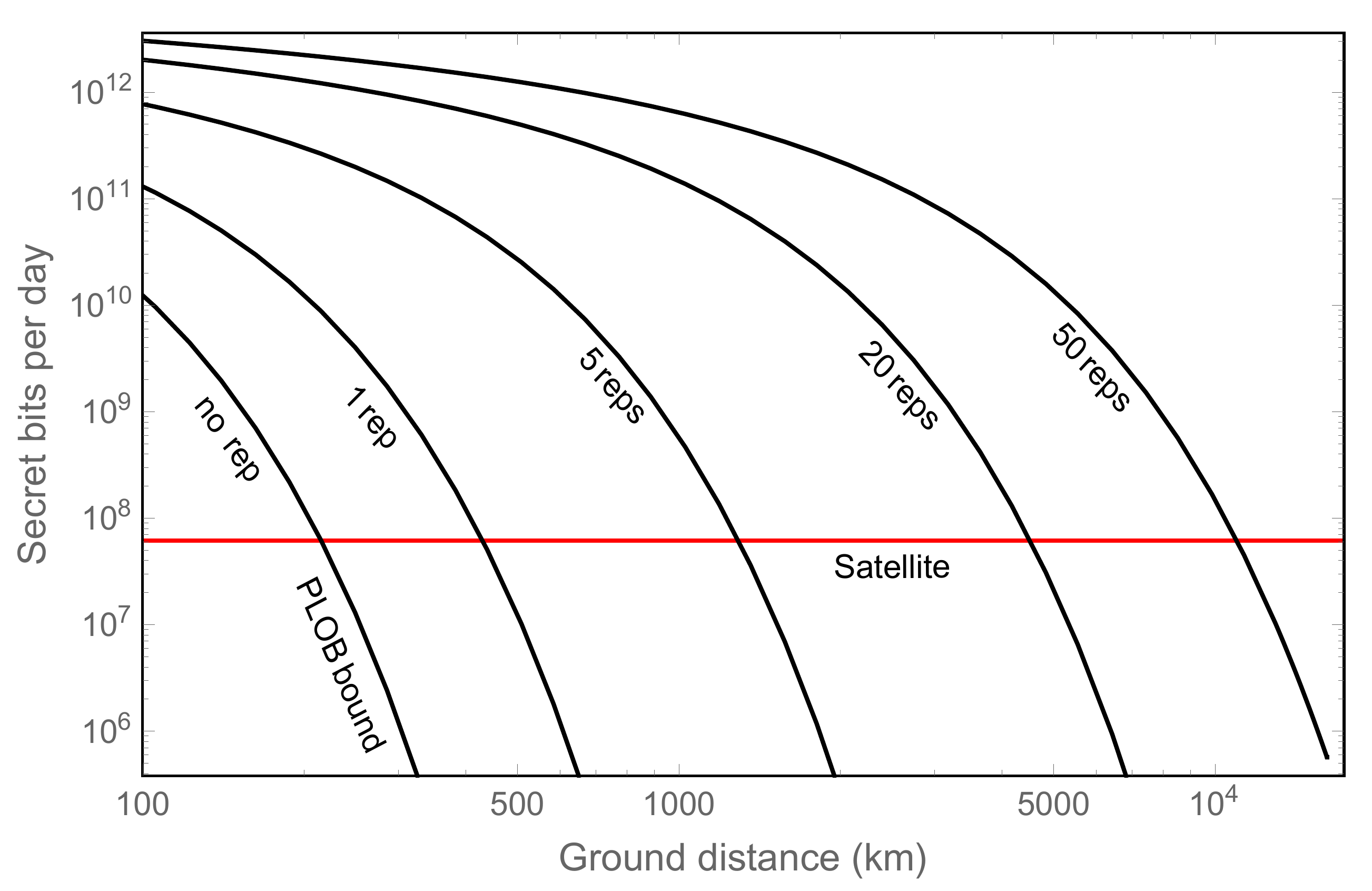}
\vspace{-0.0cm}
\caption{Comparison between ground-based and satellite-based QKD. We plot
the number of secret bits per day versus ground distance that are
distributed between two stations, assuming a clock (repetition rate) of 5 MHz. 
The black lines represent the optimal key rates achievable by a repeaterless
fiber-connection (PLOB bound~\cite{PLOB}) and by repeater-based fiber connections
assisted by 1, 5, 20 and 50 ideal quantum repeater nodes~\cite{Pirandola2019_CP}. The red line represents the 
constant key rate that is achievable by using a near-polar 
sun-synchronous satellite crossing the zenith of the two ground stations operating in trusted node. This is 
for a mean altitude of 530 km in downlink at 800 nm. The performance accounts for 
all the effects of diffraction, atmospheric extinction, (weak) turbulence, pointing error and background thermal noise. Ground-based PLOB bound is beaten at $\simeq 215$~km of standard optical fiber. See Ref.~\cite{SatBounds} for details about the parameters.}
\label{fig:sat_vs_repeaters}
\end{figure}%

Transmitting information via satellites comes with its own difficulties~\cite{Rarity2002_NJP, Aspelmeyer2003_IEEE,SatBounds,Sidhu2020,Lim2020}. Besides loss effects induced by free-space diffraction and atmospheric extinction (due to absorption and scattering), we also need to account for the inevitable beam wandering caused by pointing and tracking errors (specifically important for downlink) and atmospheric turbulence (relevant for uplink). Besides these effects, there is also the presence of background thermal noise affecting the receiver. Depending on the configuration, this may come from the sky or from the planetary albedos of the Earth and the Moon. Accounting for all these effects, one may compute the ultimate information-theoretic performances achievable by satellite quantum communications~\cite{SatBounds} together with practical key rates achievable by current technology~\cite{SatBounds,Sidhu2020,Lim2020}. In particular, one can prove that the number of secret bits that can be distributed between two remote ground stations by means of a trusted satellite can largely exceed the performance achievable by a chain of perfect quantum repeaters on the ground~\cite{SatBounds}. This is illustrated in Fig.~\ref{fig:sat_vs_repeaters}.

In designing efficient key distribution networks, it is important to assess their security against different threat models~\cite{Xu2020_RMP}. First, end-users may experience different threats to those originally envisioned by the scientists and engineers developing the technology. From a purely engineering perspective, conventional laser communications are considered more secure than radio-based systems, due to the coverage provided by the diffraction effect~\cite{vergoosen19}. It is also necessary to enquire about the necessity for the full QKD mechanism when line-of-sight is guaranteed between transmitter and receiver, using other means, such as optical beacons, radar and optical observation. In such cases, the concerns would be malicious Trojan-horse attacks, by light injection into the receiver, or harvesting scattered light into the adversary's detector~\cite{Kurtsiefer2001_JMO, Lee2019_OSA}.


\subsection{Remote clock synchronisation}
\label{subsec:clock_sync}

\noindent
Remote clock synchronisation is an important requirement to the future of quantum telecommunication, precise positioning and navigation~\cite{Sidhu2020_AVS}, and applications in fundamental science~\cite{Kolkowitz2016_PRD}. Conventional time synchronisation techniques rely on measuring the time of arrival of electromagnetic pulses~\cite{Mills1991_IEEE,Lewandowski1999_IEEE}. These classical methods are both susceptible to malicious intervention~\cite{Humphreys2008_PITMSDIN} and are limited in accuracy by the available power and bandwidth. Methods from quantum communications can address both these limitations to securely distribute high-precision time information.

Quantum clock synchronisation ensures both accurate and secure time transfer by using entanglement-based protocols~\cite{Jozsa2000_PRL, giovannetti2001quantum,Yurtsever2002_PRA}. Clock synchronisation based on arrival time of photon pairs (not necessarily entangled) can also have significantly less jitter than classical methods~\cite{ho2009clock,lee2019symmetrical}. Frequency-entangled pulses are used to construct quantum analogues of classical clock synchronisation~\cite{PhysRevD.93.065008}. Specifically, the quantum signal is used to encode both the time transfer and the secret-key generation. Once entanglement is established between optical atomic clocks, the intervening medium has no effect on the synchronisation~\cite{Jozsa2000_PRL}. This gives entanglement based clock synchronisation an additional strength. Entanglement purification operations can remove any systematic errors that arise from the use of unsynchronised clocks, which eliminate the requirement for a common phase reference between each clock within a quantum network~\cite{Okeke2018_NPJQI}. A network of optical atomic clocks operating at the fundamental precision limit was proposed in Ref.~\cite{komar2014quantum}.

The feasibility of using satellite-based quantum clock synchronisation was verified in Ref.~\cite{PhysRevD.93.065008}, which accounted for a near-Earth orbiting satellite with atmospheric dispersion cancellation. A satellite-to-ground clock synchronisation that attained a time data rate of 9 kHz and a time-transfer precision of 30 ps has been demonstrated in Ref.~\cite{dai2020towards}.


\subsection{Quantum entanglement distribution}
\label{subsec:entanglement_distr}

\noindent
Satellites and ground-based fiber quantum communication links can be used to create a global scale quantum internet. However, the potential applications of the quantum internet go beyond just QKD. Entanglement distribution enables a wide variety of other protocols including distributed and secure multi-party quantum computing, and anonymous communications protocols~\cite{Huang2020_arxiv}. Therefore, satellites must support entanglement distribution. As quantum technologies mature, higher dimensional and hyper entangled states will inevitably become of greater interest~\cite{Vallone2014_PRL}.

Currently satellites provide the most promising route entanglement distribution over global scales~\cite{Yin2017_S,Chen2021_N}. To improve the reliability of QKD services, multiple ground stations could cooperate with multiple satellites, with each link operating collectively. Namely, multiple optical ground stations (OGSs) separated by a few kilometers within a city could operate with a single satellite to deliver QKD links to end-users within the city. This increases the likelihood of link availability in the presence of intermittent cloud coverage and local turbulence effects. Furthermore, secure key store management for QKD networks with satellite links is important for improved reliability. Specifically, a secure key management ensures multiple satellite passes can be used to accumulate sufficient data for key generation~\cite{Sidhu2020}. In addition, keys will likely be used during the day but only replenished during night operation of satellite QKD where background light effects are minimised.

High-speed quantum communication with satellites is unlikely to exceed even a gigabit per second within the next 10 years, while classical communication bandwidth requirements can be several orders of magnitude higher. Planned quantum communication satellite missions can generate some few hundred megabits per year~\cite{Sidhu2020}. Thus it is impractical to use quantum secure keys in the most secure way possible -- as one-time pads where the key is the same size as the encrypted message. A compromise often implemented is to use QKD keys as seeds for classical AES encryption and update the seed with a frequency depending on the desired end to end security parameter(s)~\cite{Gheorghiu2019_arxiv}.


\subsection{Towards long range quantum entanglement networks}
\label{subsec:entanglement_networks}

\noindent
Global quantum connectivity requires multi-segment entanglement distribution links~\cite{bacsardi2018resources}. A single satellite could connect two points separated by a few thousand kilometers, with the upper limit governed by the satellites' altitude and minimum elevation through the atmosphere. A satellite in geostationary orbit could cover approximately a third of the globe. However, the achievable entanglement link rates will be heavily suppressed owing to the extreme range involved and low elevations at the extremities of the satellites trajectory. This suppression is pronounced when taking into account dual path losses for non-memory assisted entanglement distribution.

Going beyond single-satellite mediated entanglement distribution requires more complicated architectures such as entanglement swapping or quantum memories towards the realization of a space-based quantum repeater architecture. Different physical realizations have been proposed to implement a working quantum repeater depending on the loss and operational error correction techniques employed~\cite{Muralidharan2016}. In first generation QRs~\cite{Sangouard2011}, the entire communication length is divided into $2^n$ shorter segments ($L=L_0\times2^n$), where $n$ is the nesting level of the repeater. Entanglement is first created within each segment and stored in respective quantum memories (QMs) in a heralded manner. Upon the successful creation of such short-scale entangled states, the QMs are read out and the entangled state is distributed across the whole link via entanglement swapping operations. Errors originating from losses and operation are corrected through the use of heralded entanglement generation and purification protocols. This allows for the distribution of high fidelity Bell pairs across lossy channels at the expense of rate; fibre-based entanglement distribution beyond a few thousand kilometres seems unpractical even with the most ambitious schemes~\cite{Vinay2017}.

Second generation QRs use heralded generation of entanglement to correct for losses. However, they need quantum error correction to compensate for operational losses~\cite{Jiang2009}. Third generation repeaters on the other hand do not use heralding but rely fully on quantum error correction. These schemes can reach kHz rates across global distances as can be seen in Fig.~\ref{fig:sat_vs_repeaters} with a tremendous technical overhead; each individual node should contain a small scale quantum computer for error correction. In the case of third generation QRs inter-node distance must be very small so that individual channel loss is $<3$ dB ($50$\%)  in order to deterministically correct the errors. This would require placing a node every few km across the whole intercontinental link after taking coupling and detection losses~\cite{Muralidharan2014} into account.

\begin{figure}[t!]
\centering
\includegraphics[width =\columnwidth]{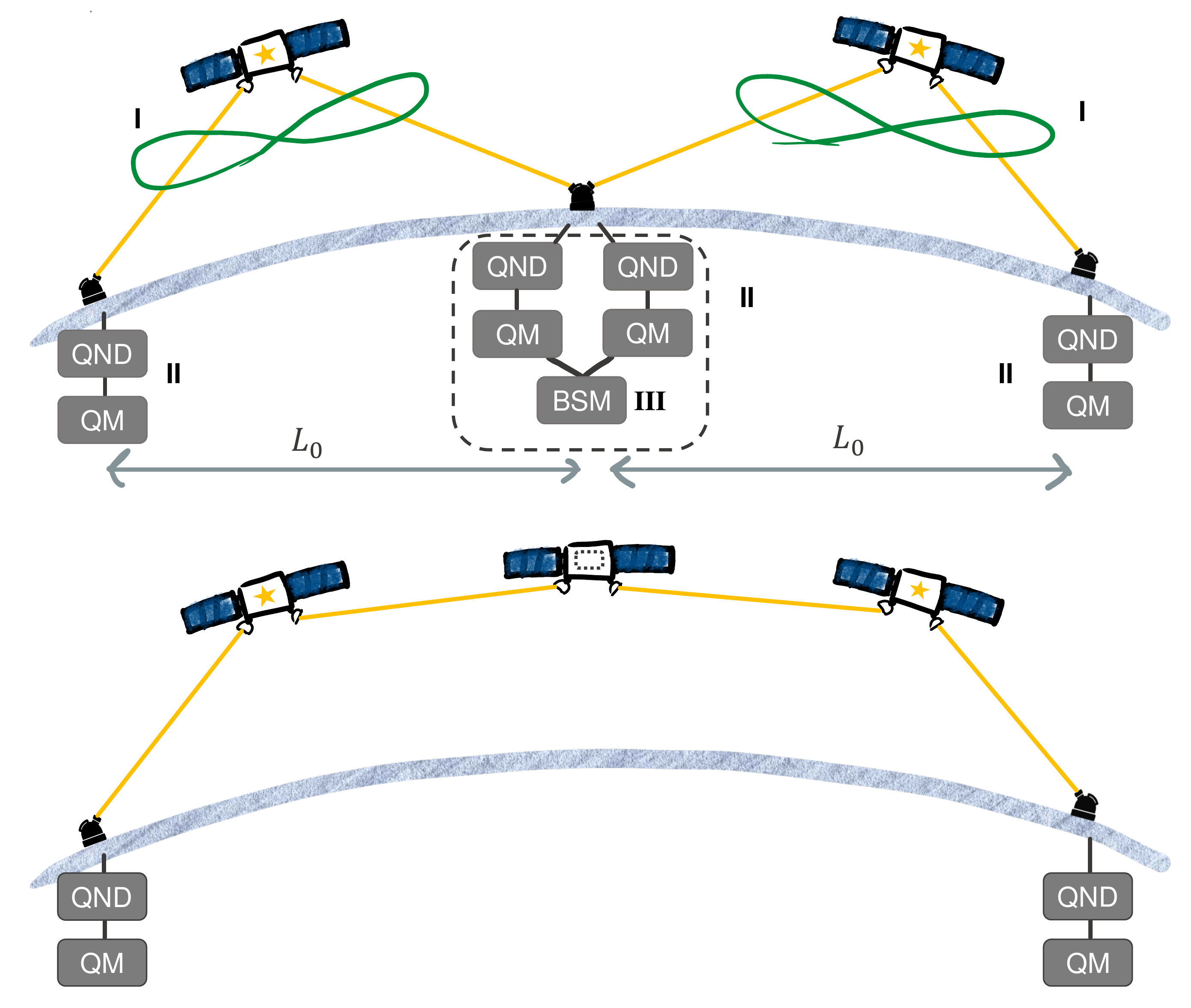}
\caption{Satellite-mediated entanglement distribution with quantum memories (nesting level $n=1$). Entangled photon pairs are generated by the sources on board satellites (step I). Quantum non-demolition (QND) measurements heralds the arrival of the photon to be stored in a memory (step II). A subsequent Bell state measurement (BSM) between different memories can then extend the range of entanglement between different end stations (step III). }
\label{fig:satellite_links}
\end{figure}

In order to extend the range of technically less demanding first generation repeaters, Boone \emph{et al.} has proposed a hybrid, space-borne QR scheme~\cite{Boone2015} where entangled photon pair sources are located on board orbiting satellites and the memories in ground stations. It was shown that meaningful key rates can be obtained across $\sim10^4$~km even with small nesting levels, $n=3$. The main limitation of this scheme is that each ground station must have good weather conditions while the link is in operation, which is extremely unlikely to achieve with $n\geq3$. A fully space-based quantum repeater scheme, such as the one in Fig.~\ref{fig:satellite_links}, have been proposed and analyzed recently~\cite{Gundogan2020_arxiv, Liorni2020_arxiv}. Placing all the components in space is expected to bring $\sim$4 orders of magnitude faster entanglement distribution rates across $\sim10^4$~km with first generation QRs (see Sec.~\ref{subsec:memories}).

A constellation of satellites equipped with entanglement sources and quantum memories will thus be required to create dynamically configurable multi-link connections between any two points on the Earth. Beyond the distribution of bipartite entanglement, it may be beneficial to create multi-partite states such as cluster~\cite{Mantri2017_SR} or graph states~\cite{Lu2007_NP}. These can be constructed from many individual entangled pairs distributed to the ground stations but it may be more efficient to develop methods by which they can be generated at the level of the satellite network. This would involve multiple quantum memories per satellite node and the ability to perform fusion operations on board the satellite. Ultimately, quantum computers may develop to the point where they can be deployed on satellites and entangled links will be needed to connect them.


\subsection{Deep-space communication}
\label{subsec:deep_space_qcomms}

\noindent
Early successes in establishing links between near-Earth satellite and ground stations increased momentum to extend satellite ranges into deep-space. Distances in this regime are generally defined in excess of 2 million kilometers via the International Telecommunications Union. For deep-space applications, the transition from radio waves to optical communications is necessary to counter low data rates that result from low channel bandwidth. It also permits improvements to both the transmitter and receiver. For the transmitter,
smaller optics are possible since optical links have a significantly narrower beam divergence. This permits the use of practically sized transmitting apertures that can reduce the size of transmitting beacons~\cite{hemmati2006deep}. For the receiver, small ground telescopes with 1-2 m diameters can be used~\cite{Czichy1995_SPIE}. Optical detection strategies also enable more capabilities than possible with radio transmission. This includes direct photo-detection, which is difficult to perform at radio photon energies. Quantum detection methods can also increase photon efficiencies for long range optical communications~\cite{banaszek2017structured}. By coherently superposing pulse position encodings, a structured receiver can interferometrically combine the received signal to produce a high peak signal measured by single photon detectors. Quantum limits to interplanetary communications was considered in~\cite{davies1977quantum}.

Despite this, communication with deep space probes remains extremely challenging compared with near-Earth optical links due to the square reduction in signal intensities with distance. Scaling near-Earth laser communication systems alone will be insufficient to overcome this challenge. Deep-space links generally require changes to the whole system design. First, on the transmitter side, higher powered lasers and transmitting optics with larger diameters will be required to reduce beam divergence. Second, on the receiver side, larger diameter optics receivers that implement efficient decoding of feint quantum signals from space using methods from quantum detection theory are required~\cite{Sidhu2020_PRXQ}. Finally, probes in deep-space have a smaller communication window that near-Earth probes, owing to an increased duration the probe orbits within near-sun angles~\cite{Hemmati2011_IEEE}.

Despite these challenges, deep-space quantum links herald a new platform to stretch experimental quantum communication experiments. These experiments include synchronisation of quantum clocks~\cite{eubanks2018time}, quantum teleportation, Bell tests, QKD, and gravitationally induced decoherence~\cite{mohageg2018deep}. The long baseline afforded by an Earth-Moon channel allows for more stringent limits to be placed upon superluminal quantum collapse propagation and free-choice loopholes in Bell tests (see Section~\ref{sec:fundamental_expts} for a discussion of fundamental tests).

\section{Space quantum communication developments}
\label{sec:current_efforts}

\noindent
The use of satellites to distribute entanglement and secure keys at intercontinental scales have been viewed as integral since the late 1990's. The first proposals for quantum key distribution from satellites to ground emerged from Los Alamos by a research team led by Richard Hughes~\cite{Hughes1999_SPIE}. Concurrently, the idea took root in Europe through an EU project involving the Defence Evaluation and Research Agency (UK) and LMU Munich. This lead to a demonstration of free space QKD over high altitude ranges with atmospheric conditions archetypal at satellite ranges~\cite{Kurtsiefer2002_N}. In addition, the European space agency (ESA) commissioned two studies on the feasibility and potential of quantum communications in space~\cite{Rarity2002_NJP, Aspelmeyer2003_IEEE}. This developed into a series of ESA studies that lead up to the Space QUEST project in 2004. A summary of these studies and the evolution of Space QUEST towards experiments on the International Space Station (ISS) were later described in Refs.~\cite{Perdigues2008_ActaAstron, ursin2009space, Joshi2018_NJP}. A key culmination of this stage of EU space research was the record breaking inter-island campaign, where a 144 km key exchange was demonstrated with both weak coherent pulse decoy states~\cite{Manderbach2007_PRL} and entanglement based sources~\cite{Ursin2007_NPhys}. 

The feasibility of space links was investigated through a series of experiments by Padua University at the Matera Laser Ranging Observatory (MLRO) of the Italian Space Agency in Matera, which was initiated in 2003. A single photon exchange from a low Earth orbit (LEO) satellite (Ajisai) to ground was realised by exploiting retroreflectors aboard the spacecraft~\cite{Villoresi2008}. This enabled the first satellite qubit transmission with retroreflectors~\cite{Vallone2015_PRL}. A small quantum bit error rate (QBER), which is defined as the ratio of the error rate to the attained key rate, was measured in this study. This provided a concrete demonstration for the practical exploitation of satellite-based quantum communications. Later studies extended the single-photon transmission distance through use of Medium Earth orbit (MEO) satellites or higher orbits, up to the current single-photon exchange limit of 20000 km~\cite{Dequal2016,Calderaro2018}.

\makeatletter
\pgfarrowsdeclare{center*}{center*}
{
  \pgfarrowsleftextend{+-.5\pgflinewidth}
  \pgfutil@tempdima=0.4pt%
  \advance\pgfutil@tempdima by.2\pgflinewidth%
  \pgfarrowsrightextend{4.5\pgfutil@tempdima}
}
{
  \pgfutil@tempdima=0.4pt%
  \advance\pgfutil@tempdima by.2\pgflinewidth%
  \pgfsetdash{}{+0pt}
  \pgfpathcircle{\pgfqpoint{4.5\pgfutil@tempdima}{0bp}}{4.5\pgfutil@tempdima}
  \pgfusepathqfillstroke
}

\pgfarrowsdeclare{centero}{centero}
{
  \pgfarrowsleftextend{+-.5\pgflinewidth}
  \pgfutil@tempdima=0.4pt%
  \advance\pgfutil@tempdima by.2\pgflinewidth%
  \pgfarrowsrightextend{4.5\pgfutil@tempdima}
}
{
  \pgfutil@tempdima=0.4pt%
  \advance\pgfutil@tempdima by.2\pgflinewidth%
  \pgfsetdash{}{+0pt}
  \pgfpathcircle{\pgfqpoint{4.5\pgfutil@tempdima}{0bp}}{4.5\pgfutil@tempdima}
  \pgfusepathqstroke
}
\makeatother

\tikzset{pics/nodetwo/.style n args={2}{
	code = {%
		\node[fill=white,opacity=40,draw=brown,rounded corners] at (#1,1) {\LARGE \textcolor{brown!80!black}{#2}};
}
}}

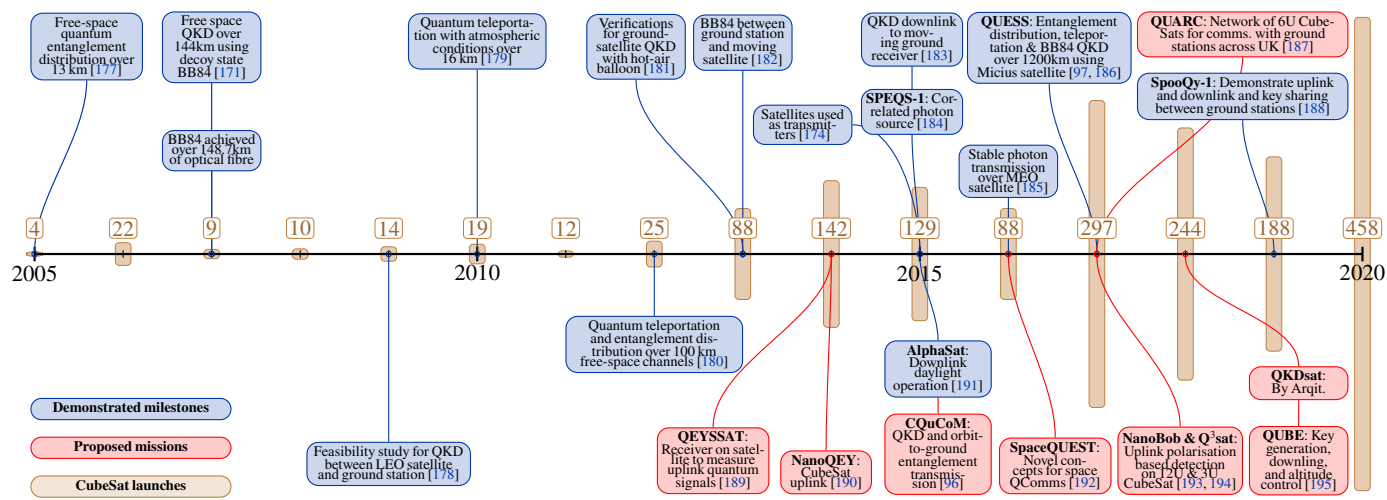
\begin{figure*}[b!]
\pgfkeys{/pgf/number format/set thousands separator={}}
\centering
        \resizebox{2.11\columnwidth}{!}{%
\begin{tikzpicture}

\def\w{0.3}
\def\scale{25} 
\def\barcolor{brown}
\def\d{17} 
\pgfmathsetmacro \unit {\d/5}
\pgfmathsetmacro \twod {2*\d}
\pgfmathsetmacro \threed {3*\d}

\filldraw[color=\barcolor,fill opacity=0.4,draw,rounded corners=.15cm] (0+\w,2/\scale) rectangle (0-\w,-2/\scale);  
\filldraw[color=\barcolor,fill opacity=0.4,rounded corners=.15cm] (\unit+\w,11/\scale) rectangle (\unit-\w,-11/\scale);  
\filldraw[color=\barcolor,fill opacity=0.4,rounded corners=.15cm] (2*\unit+\w,4.5/\scale) rectangle (2*\unit-\w,-4.5/\scale);  
\filldraw[color=\barcolor,fill opacity=0.4,rounded corners=.15cm] (3*\unit+\w,5/\scale) rectangle (3*\unit-\w,-5/\scale);  
\filldraw[color=\barcolor,fill opacity=0.4,rounded corners=.15cm] (4*\unit+\w,7/\scale) rectangle (4*\unit-\w,-7/\scale);  
\filldraw[color=\barcolor,fill opacity=0.4,rounded corners=.15cm] (5*\unit+\w,9.5/\scale) rectangle (5*\unit-\w,-9.5/\scale);  
\filldraw[color=\barcolor,fill opacity=0.4,rounded corners=.15cm] (6*\unit+\w,3/\scale) rectangle (6*\unit-\w,-3/\scale);  
\filldraw[color=\barcolor,fill opacity=0.4,rounded corners=.15cm] (7*\unit+\w,12.5/\scale) rectangle (7*\unit-\w,-12.5/\scale);  
\filldraw[color=\barcolor,fill opacity=0.4,rounded corners=.15cm] (8*\unit+\w,44/\scale) rectangle (8*\unit-\w,-44/\scale);  
\filldraw[color=\barcolor,fill opacity=0.4,rounded corners=.15cm] (9*\unit+\w,71/\scale) rectangle (9*\unit-\w,-71/\scale);  
\filldraw[color=\barcolor,fill opacity=0.4,rounded corners=.15cm] (10*\unit+\w,64.5/\scale) rectangle (10*\unit-\w,-64.5/\scale);  
\filldraw[color=\barcolor,fill opacity=0.4,rounded corners=.15cm] (11*\unit+\w,44/\scale) rectangle (11*\unit-\w,-44/\scale);  
\filldraw[color=\barcolor,fill opacity=0.4,rounded corners=.15cm] (12*\unit+\w,148.5/\scale) rectangle (12*\unit-\w,-148.5/\scale);  
\filldraw[color=\barcolor,fill opacity=0.4,rounded corners=.15cm] (13*\unit+\w,122/\scale) rectangle (13*\unit-\w,-122/\scale);  
\filldraw[color=\barcolor,fill opacity=0.4,rounded corners=.15cm] (14*\unit+\w,94/\scale) rectangle (14*\unit-\w,-94/\scale);  
\filldraw[color=\barcolor,fill opacity=0.4,rounded corners=.15cm] (15*\unit+\w,229/\scale) rectangle (15*\unit-\w,-229/\scale);  


\draw[line width=2.4pt] (0,0)--(\threed,0);
\foreach \j in {0,...,15}{\draw[black,line width=1pt] (\j*\unit,.15)--(\j*\unit,-.15);}
\foreach \i in {0,...,3}{
\draw[black,line width=2.4pt] (\i*\d,.28)--(\i*\d,-.28) node[below]{\Huge \textbf{\pgfmathparse{2005 + \i*5}\pgfmathprintnumber{\pgfmathresult}}};}


\node[fill=RoyalAzure!20,draw=RoyalAzure,line width = 1.2pt,rounded corners=.45cm,inner sep=4pt,text width=4cm,align=center] at (2,8.03) (Peng) {\LARGE Free-space quantum entanglement distribution over 13 km~\cite{Peng2005_PRL}};
\draw[-centero,line width = 1.3pt,RoyalAzure] (Peng) to[out=270,in=90,looseness=0.8]  (0,0) ;

\node[fill=RoyalAzure!20,draw=RoyalAzure,line width = 1.2pt,rounded corners=.45cm,inner sep=4pt,text width=3.5cm,align=center] at (2*\unit,8) (2006) {\LARGE Free space QKD over 144km using decoy state BB84~\cite{Manderbach2007_PRL}};
\draw[-centero,line width = 1.3pt,RoyalAzure] (2006) to[out=270,in=90,looseness=0.8]  (2*\unit,0) ;

\node[fill=RoyalAzure!20,draw=RoyalAzure,line width = 1.2pt,rounded corners=.45cm,inner sep=4pt,text width=3.5cm,align=center] at (2*\unit,4) (2007) {\LARGE BB84 achieved over 148.7km of optical fibre};
\draw[-centero,line width = 1.2pt,RoyalAzure] (2007) to[out=270,in=90,looseness=0.8]  (2*\unit,0) ;

\node[fill=RoyalAzure!20,draw=RoyalAzure,line width = 1.2pt,rounded corners=.45cm,inner sep=4pt,text width=6cm,align=center] at (4*\unit,-8.123) (Bonato) {\LARGE Feasibility study for QKD between LEO satellite and ground station~\cite{Bonato2009_NJP}};
\draw[-centero,line width = 1.2pt,RoyalAzure] (Bonato) to[out=90,in=270,looseness=0.8]  (4*\unit,0);

\node[fill=RoyalAzure!20,draw=RoyalAzure,line width = 1.2pt,rounded corners=.45cm,inner sep=4pt,text width=5cm,align=center] at (5*\unit,8.23) (Jin) {\LARGE Quantum teleportation with atmospheric conditions over 16 km~\cite{Jin2010_NP}};
\draw[-centero,line width = 1.2pt,RoyalAzure] (Jin) to[out=270,in=90,looseness=0.8]  (5*\unit,0);

\node[fill=RoyalAzure!20,draw=RoyalAzure,line width = 1.2pt,rounded corners=.45cm,inner sep=4pt,text width=6.5cm,align=center] at (7*\unit,-3.5) (Yin) {\LARGE Quantum teleportation and entanglement distribution over 100 km free-space channels~\cite{Yin2012_N}};
\draw[-centero,line width = 1.2pt,RoyalAzure] (Yin) to[out=90,in=270,looseness=0.8]  (7*\unit,0);

\node[fill=RoyalAzure!20,draw=RoyalAzure,line width = 1.2pt,rounded corners=.45cm,inner sep=4pt,text width=3.5cm,align=center] at (23.1,8.02) (Wang) {\LARGE Verifications for ground-satellite QKD with hot-air balloon~\cite{Wang2013_NP}};
\draw[-centero,line width = 1.2pt,RoyalAzure] (Wang) to[out=270,in=90,looseness=0.8]  (8*\unit,0);

\node[fill=RoyalAzure!20,draw=RoyalAzure,line width = 1.2pt,rounded corners=.45cm,inner sep=4pt,text width=3.5cm,align=center] at (27.2,8.18) (Nauerth) {\LARGE BB84 between ground station and moving satellite~\cite{Nauerth2013_NP}};
\draw[-centero,line width = 1.2pt,RoyalAzure] (Nauerth) to[out=270,in=90,looseness=0.8]  (8*\unit,0);

\node[fill=RoyalAzure!20,draw=RoyalAzure,line width = 1.2pt,rounded corners=.45cm,inner sep=4pt,text width=3.5cm,align=center] at (29.5,5) (Vallone) {\LARGE Satellites used as transmitters~\cite{Vallone2015_PRL}};
\draw[-centero,line width = 1.2pt,RoyalAzure] (Vallone) to[out=0,in=90,looseness=0.8]  (10*\unit,0) ;

\node[fill=RoyalAzure!20,draw=RoyalAzure,line width = 1.2pt,rounded corners=.45cm,inner sep=4pt,text width=3.6cm,align=center] at (33.7,8.3) (Bourgoin) {\LARGE QKD downlink to moving ground receiver~\cite{Bourgoin2015_OE}};
\draw[-centero,line width = 1.2pt,RoyalAzure] (Bourgoin) to[out=270,in=90,looseness=0.8]  (33.7,5);

\node[fill=RoyalAzure!20,draw=RoyalAzure,line width = 1.2pt,rounded corners=.45cm,inner sep=4pt,text width=3.6cm,align=center] at (33.7,5.5) (SPEQS) {\LARGE \textbf{SPEQS-1}: Correlated photon source~\cite{Ling_webpage2020}};
\draw[-centero,line width = 1.2pt,RoyalAzure] (SPEQS) to[out=270,in=90,looseness=0.8]  (10*\unit,0);

\node[fill=RoyalAzure!20,draw=RoyalAzure,line width = 1.2pt,rounded corners=.45cm,inner sep=4pt,text width=3.5cm,align=center] at (37.4,3.2) (Dequal) {\LARGE Stable photon transmission over MEO satellite~\cite{Dequal2016_PRA}};
\draw[-centero,line width = 1.2pt,RoyalAzure] (Dequal) to[out=270,in=90,looseness=0.8]  (11*\unit,0);

\node[fill=RoyalAzure!20,draw=RoyalAzure,line width = 1.2pt,rounded corners=.45cm,inner sep=4pt,text width=6cm,align=center] at (38.97,7.98) (QUESS) {\LARGE \textbf{QUESS}: Entanglement distribution, teleportation \& BB84 QKD over 1200km using Micius satellite~\cite{Yin2017_S,Liao2017_N}};
\draw[-centero,line width = 1.2pt,RoyalAzure] (QUESS) to[out=270,in=90,looseness=0.8]  (12*\unit,0);

\node[fill=red!20,draw=red,line width = 1.2pt,rounded corners=.45cm,inner sep=4pt,text width=7.8cm,align=center] at (46.4,8.5) (QUARC) {\LARGE \textbf{QUARC}: Network of 6U CubeSats for comms. with ground stations across UK~\cite{Mazzarella2020_C}};
\draw[-centero,line width = 1.2pt,red] (QUARC) to[out=270,in=90,looseness=0.8]  (12*\unit,0);

\node[fill=RoyalAzure!20,draw=RoyalAzure,line width = 1.2pt,rounded corners=.45cm,inner sep=4pt,text width=7.8cm,align=center] at (46.4,6.1) (SpooQySats) {\LARGE \textbf{SpooQy-1}: Demonstrate uplink and downlink and key sharing between ground stations~\cite{Bedington2016_EPJ}};
\draw[-centero,line width = 1.2pt,RoyalAzure] (SpooQySats) to[out=270,in=90,looseness=0.8]  (14*\unit,0);


\node[fill=red!20,draw=red,line width = 1.2pt,rounded corners=.45cm,inner sep=4pt,text width=4cm,align=center] at (26.06,-8) (QEYSSAT) {\LARGE \textbf{QEYSSAT}: Receiver on satellite to measure uplink quantum signals~\cite{Jennewein2014_SPIE}};
\draw[-centero,line width = 1.2pt,red] (QEYSSAT) to[out=90,in=270,looseness=0.8]  (9*\unit,0);

\node[fill=red!20,draw=red,line width = 1.2pt,rounded corners=.45cm,inner sep=4pt,text width=3.2cm,align=center] at (30.4,-8.45) (NanoQEY) {\LARGE \textbf{NanoQEY}: CubeSat uplink~\cite{Jennewein2014_SPIE_2}};
\draw[-centero,line width = 1.2pt,red] (NanoQEY) to[out=90,in=270,looseness=0.8]  (9*\unit,0);

\node[fill=red!20,draw=red,line width = 1.2pt,rounded corners=.45cm,inner sep=4pt,text width=3.8cm,align=center] at (34.7,-7.75) (CQuCoM) {\LARGE \textbf{CQuCoM}: QKD and orbit-to-ground entanglement transmission~\cite{Oi2017_EPJ}};
\draw[-centero,line width = 1.2pt,red] (CQuCoM) to[out=90,in=270,looseness=0.8]  (34.7,-4.45);

\node[fill=RoyalAzure!20,draw=RoyalAzure,line width = 1.2pt,rounded corners=.45cm,inner sep=4pt,text width=3.8cm,align=center] at (34.7,-4.45) (ALPHASAT) {\LARGE \textbf{AlphaSat}: Downlink daylight operation~\cite{elser2017quantum}};
\draw[-centero,line width = 1.2pt,RoyalAzure] (ALPHASAT) to[out=90,in=270,looseness=0.8]  (10*\unit,0);

\node[fill=red!20,draw=red,line width = 1.2pt,rounded corners=.45cm,inner sep=4pt,text width=4cm,align=center] at (39.2,-8.2) (SpaceQUEST) {\LARGE \textbf{SpaceQUEST}: Novel concepts for space QComms~\cite{Armengol2008_AA}};
\draw[-centero,line width = 1.2pt,red] (SpaceQUEST) to[out=90,in=270,looseness=0.8]  (11*\unit,0);

\node[fill=red!20,draw=red,line width = 1.2pt,rounded corners=.45cm,inner sep=4pt,text width=4.5cm,align=center] at (44,-8) (NanoBob) {\LARGE \textbf{NanoBob \& Q$^3$sat}: Uplink polarisation based detection on 12U \& 3U CubeSat~\cite{Kerstel2018_EPJ,neumann2018q}};
\draw[-centero,line width = 1.2pt,red] (NanoBob) to[out=90,in=270,looseness=0.8]  (12*\unit,0);

\node[fill=red!20,draw=red,line width = 1.2pt,rounded corners=.45cm,inner sep=4pt,text width=3.5cm,align=center] at (48.55,-5) (QKDsat) {\LARGE \textbf{QKDsat}: By Arqit.};
\draw[-centero,line width = 1.2pt,red] (QKDsat) to[out=90,in=270,looseness=0.8]  (13*\unit,0);

\draw[line width = 1.2pt,red] (QKDsat) to (48.55,-8);
\node[fill=red!20,draw=red,line width = 1.2pt,rounded corners=.45cm,inner sep=4pt,text width=3.5cm,align=center] at (48.55,-8) (QUBE) {\LARGE \textbf{QUBE}: Key generation, downling, and altitude control~\cite{Haber2018_Qubeproceedings}};


\pic {nodetwo={0}{\huge{4}}}; \pic {nodetwo={1*\unit}{\huge{22}}}; \pic {nodetwo={2*\unit}{\huge{9}}}; \pic {nodetwo={3*\unit}{\huge{10}}};
\pic {nodetwo={4*\unit}{\huge{14}}}; \pic {nodetwo={5*\unit}{\huge{19}}}; \pic {nodetwo={6*\unit}{\huge{12}}}; \pic {nodetwo={7*\unit}{\huge{25}}};
\pic {nodetwo={8*\unit}{\huge{88}}}; \pic {nodetwo={9*\unit}{\huge{142}}}; \pic {nodetwo={10*\unit}{\huge{129}}}; \pic {nodetwo={11*\unit}{\huge{88}}};
\pic {nodetwo={12*\unit}{\huge{297}}}; \pic {nodetwo={13*\unit}{\huge{244}}}; \pic {nodetwo={14*\unit}{\huge{188}}}; \pic {nodetwo={15*\unit}{\huge{458}}};

\begin{scope}[xshift=2.5cm, yshift=0.0cm]
\node[fill=RoyalAzure!20,draw=RoyalAzure,line width = 1.2pt,rounded corners=.4cm,inner sep=4pt,text width=7.4cm,align=center, text height = 0.4cm,text depth = 0.14 cm] at (1.2,-6) {\textbf{\LARGE Demonstrated milestones}};

\node[fill=red!20,draw=red,line width = 1.2pt,rounded corners=.4cm,inner sep=4pt,text width=7.4cm,align=center, text height = 0.4cm,text depth = 0.14 cm] at (1.2,-7.5){\textbf{\LARGE Proposed missions}};

\node[fill=\barcolor!20,draw=\barcolor,line width = 1.2pt,rounded corners=.4cm,inner sep=4pt,text width=7.4cm,align=center, text height = 0.4cm,text depth = 0.14 cm] at (1.2,-9){\textbf{\LARGE CubeSat launches}};
\end{scope}

\end{tikzpicture}%
}
\caption{Timeline of key milestones in field demonstration and feasibility studies towards the developments of satellite-based QKD. This includes ground tests and establishing entanglement links with orbiting satellites. These milestones are indicated with blue boxes. Red boxes indicate proposed missions and their objectives. The number of CubeSat launches are illustrated in brown~\cite{Kulu2020_database}. Notice the increase in the number of missions involving CubeSats reflects their growing importance in satellite-based global quantum communications.}
\label{fig:satellite_mission_timeline}
\end{figure*}

The use of optical corner reflectors or corner cubes improves the OGSs ability to acquire and track the satellite. The improved accuracy in pointing, acquisition, and tracking (PAT) enabled studies of the superposition principle with temporal modes, i.e. time-bin interference~\cite{vallone2016}. By combining the temporal and polarisation degrees of freedom of a single photon, the wave-particle duality was then tested in space following the John Wheeler gedanken experiment of the delayed-choice~\cite{vedovato2017}. This experiment confirmed the quantum mechanical prediction in a novel and much larger scale compared with ground tests.

Between 2008~\cite{armengol2008quantum} and 2018~\cite{Joshi2018_NJP}, EU teams showed great progress in ground demonstrations and technology platform developments, though without attracting additional support to reach mission threshold. Meanwhile groups in China have taken up the mantle launching the world’s first major quantum communications satellite as part of a much larger quantum technologies program~\cite{Liao2017_N, Yin2017_S, Ren2017_N}. Stimulated by these outstanding results, several groups world-wide have now made rapid progress towards satellite quantum communications missions, many of which are reviewed in this paper. Groups from Canada, Japan, Singapore, Switzerland and the UK have started to look at commercial exploitation of quantum communications missions.

Fig.~\ref{fig:satellite_mission_timeline} illustrates a timeline of key missions that have demonstrated key milestones or have conducted feasibility studies of global, satellite-based quantum communications. Proposed missions and their scientific objectives are included. We also illustrate the number missions that take advantage of of small satellite. The dramatic increase demonstrates the importance of small satellites in future quantum space missions. For a complete exposition of satellite missions, the reader is directed to Refs.~\cite{Bedington2017_npjQI,Lee2019_arxiv}.


\subsection{QUESS satellite}
\label{subsec:quess_satellite}

\noindent
Quantum Experiments at Space Scale (QUESS) is a Chinese research project operated by the Chinese Academy of Sciences (CAS). The mission uses a LEO satellite called Micius, to demonstrate space-based quantum communication. This successfully demonstrated integration with existing ground-based networks, generating unconditionally secure quantum cryptographic keys over intercontinental distances between Asia and Europe, and the demonstration of quantum entanglement distribution and quantum teleportation at space scales.

To date, this initiative has demonstrated three key milestones towards a global-scale quantum internet. First, it has achieved secure satellite-to-ground exchange of cryptographic keys between the Micius satellite and multiple ground stations in China. It has implemented decoy-state QKD with a kilohertz key rate over a distance of 1200 km~\cite{Liao2017_N}. By using a weak coherent pulse at high channel losses, the keys are secure because photon-number-splitting eavesdropping can be detected. This key rate is around 20 orders of magnitudes greater than that expected using an optical fibre of the same length. Second, the Micius satellite has demonstrated the capability of two-photon entanglement distribution to ground stations separated by ~1200 km, and a violation of Bell inequality of $2.37 \pm 0.09$ under strict Einstein locality conditions~\cite{Yin2017_S}. After significant improvements in collection optics which improved the link transmission, sufficient number of photons can be detected to realize the BBM92 protocol across 1120~km~\cite{Yin2020_N}. This is a crucial stage to demonstrate the feasibility of a true global quantum communication network. Third, quantum teleportation of independent single-photon qubits has been demonstrated through an uplink channel for ground-to-satellite quantum teleportation over distances of up to 1400 km~\cite{Ren2017_N}. This demonstration successfully teleported six input states in mutually unbiased bases with an average fidelity of $0.80~\pm~0.01$, which is above the optimal state-estimation fidelity on a single copy of a qubit~\cite{Massar1995_PRL}. Improving this fidelity is an essential future step to enabling space-based quantum repeaters.

The Micius satellite has also be used as a trusted relay between different ground stations for high-security key exchange. This was demonstrated in a video conference between China and Vienna for intercontinental quantum-secured communication~\cite{Liao2018_PRL}. In addition, a large-scale, hybrid quantum communication network has been realized by integrating space links provided by the Micius satellite to an already existing 2000~km long Beijing-Shanghai trusted node link resulting in a total quantum communication distance of 4600~km~\cite{Chen2021_N}. This work is the first example of an intercontinental scale QKD network with around 150 users. These experiments are the first steps toward a global space-based quantum internet. 


\subsection{Small satellite efforts}
\label{subsec:small_Satellites}

\begin{figure}[t!]
\centering
\includegraphics[width =0.95\columnwidth]{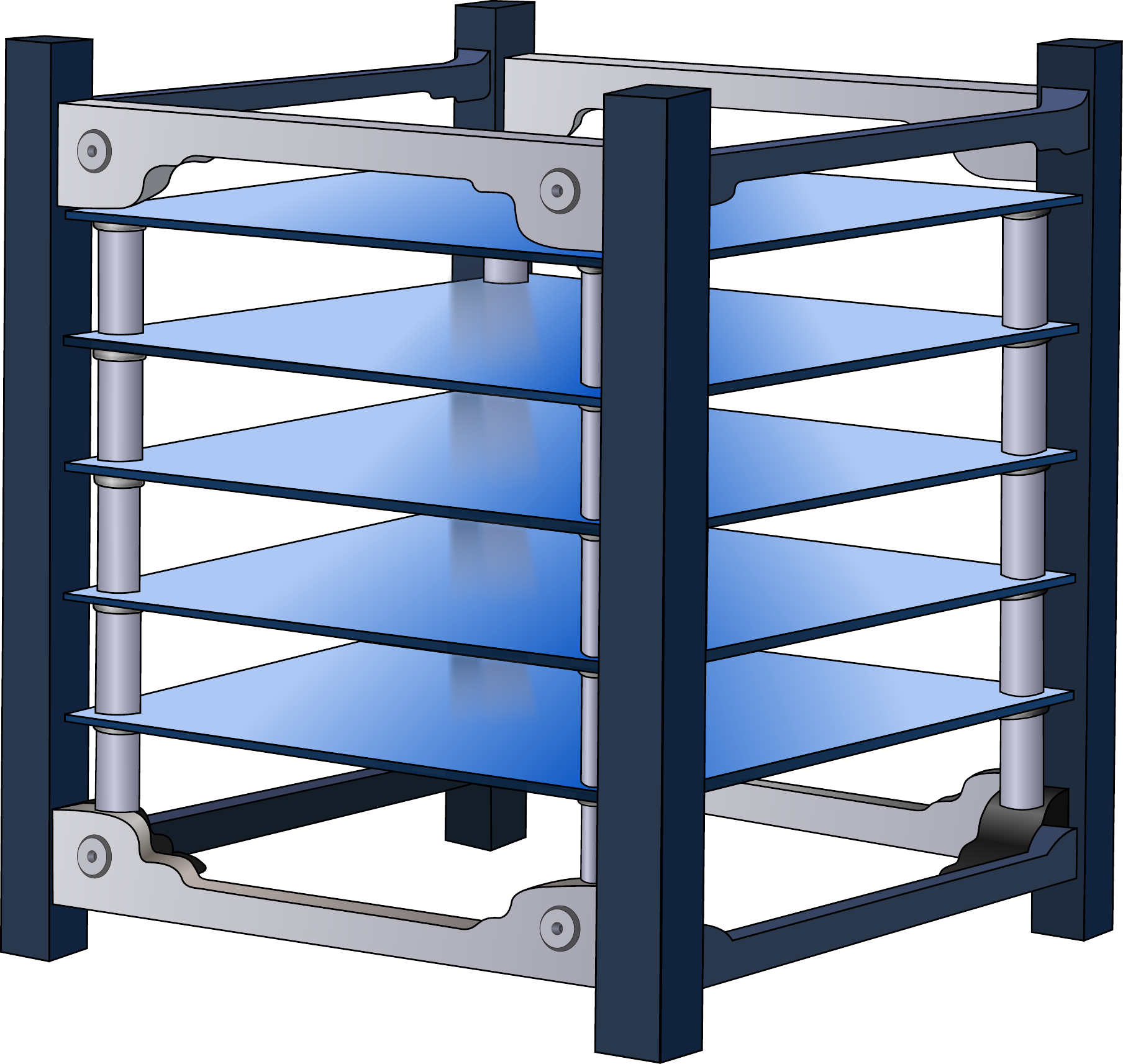}
\caption{An illustration of a 1-unit CubeSat chassis with dimension 10 cubic cm based on the CubeSat design specification standard. Miniaturised satellites can be assemble by loading each frame with the necessary components, such as power systems, communications, altitude determination and control systems, control. Each unit can be readily stacked to form a multiple unit, modular nanosatellite. This provides a low-cost and time efficient solution for quantum communication implementations in space.}
\label{fig:CubeSat_chassis}
\end{figure}

\noindent
To demonstrate and implement a global quantum communication network with multiple users and reasonable coverage we require a constellation of satellites. This presents a significant obstacle for the Micius initiative. Specifically, large satellites are expensive to develop and will take substantially more time to develop such a constellation of low-Earth orbit trusted-nodes for QKD service provision. An alternative approach that has received a surge of interest is the use of CubeSats~\cite{Oi2017_CP}. These are miniaturised nanosatellites for space research that are made up of module units of $10 \text{cm} \times 10 \text{cm} \times 10 \text{cm}$ cubic units, and a mass of no more than 1.33 kilograms per unit. The chassis of a single unit CubeSat is illustrated in Fig.~\ref{fig:CubeSat_chassis}. The miniaturised size of CubeSats limits the size of onboard optical telescope apertures, the volumetric space, weight, and power (SWaP) required by the payload, thermal design, and pointing stability of the platform. Together with their lower development cost, shorter development times, and increased deployment opportunities, CubeSats have the potential to deliver rapid progress in space quantum technologies that are expected to surpass conventional space systems development~\cite{Tang2016_PRApp, Tang2016_SR,Villar2020_O}. This has ushered in the CubeSat era in space research~\cite{Latt2014_PEAS, Wuerl2015_IEEE}. 

The 6U CubeSat platform is commonly used since it delivers the largest size with a favourable cost to capability trade-off for many high-performance nanosatellite missions~\cite{Tsitas2012_AJ}. Several design studies have used 6U CubeSats for Earth observation as it accommodates a reasonably large optical assembly together with ancillary payloads. There are approximately 65 6U missions under development. Since CubeSats are not restricted to Earth orbits, some missions choose the relatively bulky 12U form factor to be able to accommodate the largest possible telescope aperture on a CubeSat~\cite{Kerstel2018_EPJ}. A review on all current approaches of satellite QKD enabling initiatives has been summarised by Bedington~\emph{et al.}~\cite{Bedington2017_npjQI}. An advantage of the CubeSat approach is the availability of conventional off-the-shelf components in order to reduce costs and development time. We briefly summarise a selection of small satellite mission proposals, development, and upcoming launches in the remainder of this section.

\subsubsection{QEYSSat}
\label{subsubsec:qeyssat_satellite}

\noindent
The Canadian Quantum Encryption and Science Satellite (QEYSSat) is a funded mission for a micro satellite in LEO with expected completion in early 2022~\cite{CSA_QEYSSat}. The primary objective is to study and advance the science of quantum communication links and technologies, as to demonstrate long-range quantum entanglement~\cite{Rideout2012_IOP}. QEYSSat is viewed as a precursor to larger scale missions and quantum networks. 

The mission also anticipates to explore the use of different quantum sources on the ground, and transmit them to a receiver aboard the microsatellite~\cite{ISI:000335950400004}. While delivering the photons reliably to the moving satellite via an optical up-link channel is technically more challenging than a down-link configuration, it offers greater flexibility to study quantum links with various quantum sources and quantum interfaces introduced at the ground segment. For instance, interfacing fibre-optical networks with the satellite link can be accommodated by a dual-wavelength entangled photon source where one photon wavelength matches the space link ($\approx 785$~nm) and the entangled partner photon's wavelength matches the fibre optical link ($\approx 1550$~nm). 


\subsubsection{CQuCoM}
\label{subsubsec:cqucom}

\noindent
Growing out of bilateral effort between the UK and Singapore on developing space quantum technologies~\cite{morong2012quantum,bedington2015deploying,Tang2016_PRApp,Bedington2016_EPJ}, the CubeSat Quantum Communications Mission (CQuCoM) was proposed by a 6-nation consortium in 2015~\cite{Oi2017_EPJ}. The mission concept comprised a 6U CubeSat that would transmit quantum signals to Earth, with the Matera Laser Ranging Observatory in Italy being the primary receiving ground station. This would help establish efficient constellations of low-Earth orbit trusted-nodes for QKD service provision. In addition, the mission would include atmospheric visibility simulation, and satellite QKD threat analysis. The project leveraged the lower costs of the platform and orbital ride-share to propose two successive launches. The first mission would implement the decoy-state weak coherent pulse BB84 protocol as a low risk pathfinder for the second, that will implement QKD with entangled photon sources. The CubeSats would have been deployed from the ISS into a 400 km circular orbit with a projected mission lifetime of 12 months. The programme would have exploited the rapidly improving nanosatellite altitude determination and control systems (ADCS) as demonstrated on contemporary missions such as the MinXSS~\cite{mason2017minxss} and the bright target explorer (BRITE)~\cite{pablo2016brite}. CQuCoM inspired several follow-on CubeSat missions, including SpeQtre, QUARC/ROKS, QUBE, NANOBOB/Q$^3$sat, which are described in the following sections.


\subsubsection{SpeQtre}
\label{subsubsec:qkdqubesat}

\noindent
This project is another bilateral UK-Singapore effort coordinated by RAL Space and the Centre for Quantum Technologies~\cite{dalibot2020speqtre}. In this effort the goal is to demonstrate entanglement distribution between a CubeSat platform and optical ground receivers on Earth. This would pave the way for a constellation of entanglement transmitters from space, and is under active development. This project draws heritage from previous Singaporean quantum CubeSat missions launched in 2014, 2015, and 2019~\cite{Lim2020_arxiv}. RALSpace directs the development of the spacecraft bus capable of establishing the optical link for entanglement distribution, and the design, development, and implementation of the optical ground station. One of the objectives in this mission is to test optical ground receivers near major population centres, to implement connectivity with existing metropolitan quantum networks. 

This program has also studied the effects of different constellations on entanglement distribution across the globe. This included studying different constellation arrangements for serving the Indo-ASEAN region, which concluded that over the course of a single year, tens of Mbits of secret keys could be distributed after consideration of atmospheric effects~\cite{Vergoossen2020_AA}.

\subsubsection{QUARC}
\label{subsubsec:quarc}

\noindent
The quantum research CubeSat~(QUARC) is a UK initiative for QKD service provision across the UK. The initial scope was concentrated on miniaturised pointing and tracking subsystems suitable for CubeSat payloads. Since, the mission has expanded to the design and construction of compact decoy-state weak coherent pulse BB84 sources and a portable optical ground station. A preliminary use case was the security of UK critical national infrastructure~\cite{polnik2020scheduling} with a detailed study of link availability based on weather patterns, and identifying availability of suitable OGS locations close to metropolis regions. Tradeoffs between. For a 15 CubeSat constellation topology servicing 43 ground stations distributed uniformly across the UK, trade-offs between the topology, atmospheric channel, and achievable key rates were considered in Ref.~\cite{mazzarella2020quarc}. 

The best location for a fixed QKD OGS is one which is close to the end users (most likely in urban centers) but sufficiently far enough away to avoid the worst of light pollution. A mobile OGS will be able to identify and trial such sites. 
It will consist of a 43\,cm telescope system with attached modules for processing the quantum signal, timing and synchronisation, tracking and fine pointing and the beacon systems all mounted in a vehicle~\cite{Zhang2020_conference}. The ability to rapidly move and deploy the OGS may help overcome regional cloud cover limitations.


\subsubsection{UK QT Hub mission}
\label{subsec:UK_QT_missions}

\noindent
As part of the UK National Quantum Technology Programme, the quantum technology hub in quantum communications is planning to launch a CubeSat mission to demonstrate space to ground QKD in the 2023-24 timeframe~\cite{qthubspacesector}. This will leverage design specifications from the previous SpeQtre mission and the RAL Space developed 12U CubeSat platform to host a UK-developed source. A UK OGS facility will be developed to support the mission. Lineage from the QUARC programme will feed into these developments. 


\subsubsection{QUBE}
\label{subsec:QUBE}

\noindent
This is a German 3U satellite that will implement a downlink of strongly attenuated light pulses for exchange of encryption keys. The goal of this first phase of the project is to test the components and the viability of performing QKD between a single CubeSat in LEO and an optical ground station OGS in Germany~\cite{Haber2018_Qubeproceedings}.


\subsubsection{Q$^3$sat}
\label{subsubsec:q3sat}

\noindent
The Quantum-cubed satellite is an Austrian satellite being developed as a compact 3U uplink quantum satellite~\cite{neumann2018q}. The mission aims to produce low cost secure keys at $\approx$ EUR20 per kbit. This minimalist mission uses only two single photon detectors with minimal radiation shielding and a slow active basis choice using liquid crystal polarisation modulators. Further, its limited battery and solar panel capacity only allow one active downlink every two orbits. Nevertheless, its low cost makes this the ideal option for early adaptors of the technology.


\subsubsection{NanoBob}
\label{subsubsec:nanobob}

\noindent
The NanoBob~\cite{Kerstel2018_EPJ} is a 12U French-Austrian initiative to create a versatile QKD receiver in space. The 12U form factor was chosen to accommodate the largest possible telescope aperture in a standard CubeSat. Using the XACT-15 ADCS manufactured by Blue Canyon Technologies, its body pointing is estimated to be precise and fast enough, to track a grond station or another QKD satellite (in a suitable orbit) without a fine pointing mechanism. It utilises a passive polarisation basis choice and four single-photon detectors. A cost-benefit analysis of different orbits concluded that a Sun Synchronous Orbit would provide the best key rates but a circular orbit has much lower launch costs and would be the more economical choice.


\subsection{NASA projects}
\label{subsec:Nasa}

\noindent
NASA is pursuing multiple mission studies that aim to develop a global
quantum communications infrastructure through a space-based
distribution of quantum entanglement \cite{NasaSpaceQuantum,NSTCreport}  A LEO platform (such as the International Space Station) \cite{wiredquantum} 
could distribute entanglement between ground stations as far as
1200 km apart --similar to the Micius accomplishments.  A similarly outfitted spacecraft in a higher orbit
could connect ground stations separated by longer baselines.  The spacecraft architecture may consist of a high clock-rate
source of high-fidelity entangled photon pairs, a detector array
that performs quantum state tomography on the photon pairs, and
a dual-telescope gimbal system. 
Furthermore, to overcome the high losses associated with an untrusted node scenario, high
clock-rate sources, exceptional pointing accuracy in all telescopes,
high photon throughput on both the transmitter and receiver ends,
and large, diffraction-limited apertures (possibly aided by adaptive
optics) will be required. 
 See section~\ref{sec2:challenges} for further discussions.

The Deep Space Quantum Link (DSQL) \cite{2018LPICo2063.3039M,MazzarellaOSA20} is a NASA project that
plans to perform pioneering experiments to explore relativistic effects
on quantum systems. DSQL plans to access the regime where the effects of
special and general relativity affect the outcome of uniquely
quantum primitives such as teleportation, entanglement distribution, and the violation
of the Bell’s inequalities. These effects have a more pronounced experimental impact over long-baseline channels between exotic orbits, e.g. retrograde and highly elliptical.  For example, one potential orbital deployment for DSQL is
the Lunar Gateway
~\cite{Rideout2012_IOP, spann2017science}, a space station orbiting the moon
that will establish a quantum link with Earth-based ground stations
or high-altitude platforms orbiting the Earth.

The technology development required for DSQL mirrors the
requirements for high-performance quantum communications systems. In particular,
support of Lunar-Earth links needs a minimum pair production rate of the
order of 100 MHz to overcome the high loss channel and deliver more photon counts than background events. Such a figure corresponds to a clock
rate of 10 GHz or higher, assuming a 1\% pair production probability
to produce high-fidelity photon pairs. The single-photon detector
system would have to operate at a high-rate and efficiency and with
low jitter and background noise. For example, to perform the optical test of the Einstein Equivalence Principle proposed in \cite{ Terno2018a}, exceptionally low jitter is required. DSQL could as well enable or improved some experiments by using quantum memories.


\subsection{Other initiatives}
\label{subsec:other}

\noindent
An even simpler approach to quantum space science, pioneered by a team at the University of Padua in Italy in collaboration with Italian Space Agency, involves adding reflectors and other simple equipment to regular satellites. 
By exploiting corner-cube retroreflectors (CCR) on board of already orbiting satellites currently used for satellite laser ranging, it has been possible to demonstrate the single photon exchange from a LEO satellite to the Matera Laser Ranging Observatory ground  station~\cite{Villoresi2008, Vallone2015_PRL}. 
After these first demonstrations, the technique has been used to extend the single photon transmission up to an MEO satellite~\cite{Dequal2016} and more recently to GNSS orbit~\cite{Calderaro2018}.
Moreover, the team showed that photons bounced back to Earth off an existing satellite maintained their quantum states and were received with low enough error rates for quantum cryptography~\cite{Vallone2015_PRL}. Through this method, it could be possible to generate secret keys, by replacing passive CCR with a modulated version \cite{Vallone2015_PRL,Rabinovich2018}. The advantage of such a scheme is that it does not require precise pointing of the spacecraft toward the ground station. On the other hand,  velocity aberration limits the diffraction pattern of the retroreflected pulse, thus imposing a non-optimal solution for the beam transmission efficiency. The CCR technique has been also adopted by a Chinese collaboration to test single photon exchange back-reflected from the de-orbiting satellite CHAMP~\cite{Yin2013}.

Besides CCR, other existing satellites have been used to perform proof-of-principle demonstrations of satellite quantum communications. This is the case of the geostationary satellite Alphasat I-XL, whose optical payload LCT has been used to study the propagation of CV encoded quantum states beyond Earth atmosphere. A German collaboration was able to demonstrate the quantum limited detection of coherent states of light transmitted by LCT payload and received by the Transportable Adaptive Optical Ground Station, located at the Teide Observatory~\cite{Gnthner2017}. The measurement demonstrated the persistence of coherent properties of light from satellite also in the quantum limit and it was able to assess the excess noise on such a communication channel.


\subsection{Commercial efforts}
\label{subsec:commercial_efforts}

\noindent
There are numerous planned commercial QKD missions. First, the European Space Agency (ESA) is supporting two large commercial efforts: QKDSat and QUARTZ. These are led by Arqit and SES respectively, and form part of the ARTES Secure and Laser communication Technology (ScyLight) programme~\cite{toyoshima2020recent}. ESA is also working on the security and cryptographic mission (SAGA), which aims to European-wide quantum network with space segments~\cite{qciwhitepaper2019}. Second, Craft Prospect aims to demonstrate in-orbit autonomous operations for key transfer using machine learning techniques by 2022~\cite{mercury2020quantum}. This effort is part of the responsive operations and key services (ROKS) mission, which is a continuation of the QUARC programme that uses a 6U CubeSat platform. Finally, SpeQtral Pte Ltd. is a spin-out company in Singapore that aims to demonstrate entanglement distribution using CubeSats~\cite{Tang2016_PRApp,Villar2020_O} in the upcoming SpeQtre mission~\cite{dalibot2020speqtre}.



\section{Space quantum communication challenges}
\label{sec2:challenges}

\noindent
Satellite-based quantum technologies allows quantum communication over global ranges.  However, there is a number of challenges that must be overcome. First, the propagation of quantum signals through free-space is subject to several mitigating sources of noise. These include pointing errors, geometric diffraction, atmospheric turbulence, and background noise from stray light and daylight operation. Second, quantum technologies must endure the demanding conditions of space. Engineering each component for space preparedness and establishing a networked satellite infrastructure presents unique challenges. In this section, we provide an overview of some of these major challenges and highlight current efforts that aim to mitigate their effects.


\subsection{Protocols, system performance, and optimisation}
\label{subsec:protocols}

\noindent
Operating satellite QKD has several additional complications over terrestrial implementations. Specifically, there is a restricted transmission time owing to a satellite overpass and a highly variable channel loss both within and between pass. A reduced transmission time imposes finite block effects that limit the attainable secret key~\cite{Sidhu2020}. This contrasts with optical fibre based systems where the protocol proceeds until a predetermined (typically large) block of data is acquired. Additionally, stochastic effects of the atmosphere directly improved studies on channel losses. Current modelling and analyses on the performance of QKD have focused on terrestrial-based fibre channels. The analyses of currently existing protocols must be adapted if we are to understand their performances for satellite QKD. This is important both to aid understanding and help guide the design and improvements for future missions.

In addition, the optimisation of system parameters, such as the basis encoding probabilities and signal intensities for decoy state implementations is an important feature for key generation under finite block effects. This optimisation may be constrained owing to physical limitations to some parameters. Understanding the effects of different system parameters on the attainable key has been studied in Ref.~\cite{Sidhu2020}.


\subsection{System dependent space-Earth channel link losses}
\label{subsec:loss_sources}

\noindent
The link between the satellite and OGS is characterised by the high losses (or low transmissivity) that the quantum signal experiences. Numerical studies of the optical channel between space and ground predict a link loss of 30-40 dB for a spacecraft with a 10 cm aperture at 500 km altitude and a 1 m aperture at the optical ground station~\cite{Bourgoin2015_OE}. Micius achieved a total system link loss of 27 dB with 18-30 cm apertures (non-diffraction limited) and 1.2 m diameter ground receivers~\cite{Yin2020_N}. The largest contributor to this loss is the divergence of the beam due to diffraction, which scales as the inverse square of the propagation distance with the final beam spot size typically several times larger than the receiving telescope.

Other system factors can increase loss including internal transmitter and receiver inefficiencies, non-ideal photon detection efficiency, and pointing inaccuracy. Due to the narrow optical beam divergence necessary for minimising diffraction spread, the pointing of the telescope has to be controlled to an extremely precise degree, of the order a microradian. The satellite must be able to determine its attitude and position with respect to the OGS, typically with star trackers and Global Position System (GPS) receivers, and then control its direction using mechanisms such as reaction wheels and magnetic torque coils. A body-mounted telescope can thus be coarsely pointed towards the OGS position, so-called latitude-longitude-altitude (LLA) pointing. For small satellites, open-loop LLA pointing accuracy of 0.01 degrees has been demonstrated~\cite{rose2019optical} hence an additional mechanism can be employed using a fast steering mirror (FSM) to provide the microrad fine pointing required. For pointing error a fraction of the beam divergence, the additional loss can be considered minor.

\subsection{Impact of the atmosphere: turbulence and noise}

\noindent
A quantum signal propagating through the Earth’s atmosphere is affected by turbulence and an increase in background noise owing to daylight operation. Both effects directly limit the attainable key and entanglement distribution rates. The effects of turbulence are well understood from classical optics~\cite{Kaushal2017_book}. Specifically, interaction with turbulent eddies in the atmosphere cause rapid fluctuations in the transmission efficiency of a communication link. This leads to a stochastic broadening of the beam waist and wandering of the beam centroid~\cite{Yura1973_JOSA}. Turbulence effects on DV and CV QKD have also been well studied with ground-to-ground horizontal links~\cite{FreeSpaceBounds,Vallone2015_PRA, Heim2014, Shen2019}. An uplink configuration where the quantum signal is sent from the ground to space, will have significantly more loss due to the increased beam wander caused by atmospheric turbulence than a downlink~\cite{vasylyev2019satellite}.

Generally, the sensitivity of the quantum signal to atmospheric turbulence depends on the type of encoding chosen. For example, information in CV QKD protocols is encoded in the amplitude of Gaussian quadrature-modulated coherent or squeezed states of light~\cite{QKDreview2020}. These signal fluctuations are directly affected by atmospheric turbulence and weather conditions~\cite{Vasylyev2017_PRA,Ruppert2019}. These transmittance fluctuations are known as channel fading and their impact on CV QKD have been well studied~\cite{FreeSpaceBounds,Usenko2012_NJP, Heim2014, Papanastasiou2018_PRA}. 

Fortunately, there are some methods to enhance robustness against these effects. Hybrid polarisation-spatial mode encoding techniques can be more resilient against atmospheric turbulence in some circumstances than pure orbital angular momentum (OAM) states~\cite{farias2015resilience}. Also, adaptive optics can correct for the effects of atmospheric turbulence in QKD systems~\cite{wang2019performance,pugh2020adaptive} and become increasingly effective for large telescope diameters, with typical loss reduction by as much as 3 to 7\,dB. Adaptive optics are most beneficial for systems that need single mode coupling. The main drawback of adaptive strategies is that establishing a suitable reference beam for closed loop feedback may not always be practical. An alternative is to use a 4-f optical imaging system to ensure near perfect mode overlap despite significant beam wander. This has been used to demonstrate free-space time-bin QKD, where interferometric measurements are necessary~\cite{jin2019genuine}.

Daylight operation of QKD leads to increased back-scattering of solar radiation in the atmosphere, which increases the background noise in the signal. The effects of daylight operations on QKD has been investigated for more than two decades and several demonstrations of free space QKD in daylight have been realised~\cite{Buttler2000, Hughes2002, Peloso2009}. More recently, by exploiting single mode fibre coupling as spatial filtering and by using telecommunication wavelength to minimise the external noise, it has become possible to demonstrate QKD in daylight with the potential to reach satellite to ground communication~\cite{Liao2017, avesani2019daylight}. Previous demonstrations used near-IR, narrow wavelength filtering and baffles around the receiving telescope~\cite{Peloso2009}. In particular, in ~\cite{avesani2019daylight}, the authors  proposed an integrated photonic circuit as a source of quantum states. This is an interesting technology for space application, as it might drastically reduce the size, weight and power (SWaP) of the payload. CV QKD protocols may offer advantages in terms of reducing the amount of external background noise collected by the receiver by exploiting the mode-matching of the signals with the bright local oscillator in homodyne-like setups, so that the local oscillator (transmitted or locally generated) acts as an effective and natural noise filter in frequency~\cite{FreeSpaceBounds}. This paves the way for potential implementations of CV QKD protocols in day-light conditions for both downlink and uplink~\cite{SatBounds}.


\subsection{SWaP for space segment}
\label{subsec:SWAP_criteria}

\noindent
The physical size of a quantum communication devices are limited by their complexity, size of constituent components for bulk optical setups, and waveguide sizes for photonic chip scale setups. Reducing the total satellite weight is a mechanical optimisation, since designers must ensure that the satellite is sufficiently robust to survive launch. Radiation shielding, reaction wheels and, for larger satellites, the satellite structure itself can also increase satellite mass, leading to increased launch cost. Higher electrical power consumption can be accommodated by deployable solar panels, though at the cost of greater complexity, risk, larger electrical power subsystem, and thermal control requirements. Else the duty cycle of the payload may need to be restricted to fit within the orbit averaged power limit. Active temperature stabilisation, high powered beacon lasers, intensive computation and the ADCS (both reaction wheels and magneto-torquers) are typically power hungry subsystems~\cite{NASASOTASST}.

From sender to receiver, Terrestrial QKD systems are often bulky and expensive devices, for example a Clavis3 device from idQuantique~\cite{clavis3} has a volume of 24.5 litres and weighs approximately 10 kg. While this is not an issue for applications in network infrastructure, this presents significant issues in other applications. There exist smaller QKD transmitters ($\approx100 \text{cm}^3$) \cite{Vest2015, Chun2017,zhou2019polarization} but these do not address the size, weight, and power (SWaP) of the receiver section, and so limit the available topologies for implementations. None of the systems cited above were designed for space so there can be some conclusions reached from examining them however focusing on systems designed for space, the lowest SWaP systems with space heritage are SOTA ($1000 \text{cm}^3$, 6.2 kg, 39.5W)~\cite{carrascocasado2017} and SPEQS ($330 \text{cm}^3$, 300 g, 2 W)~\cite{chandrasekara15}.

A promising solution to the SWaP of the payload is to use optical chip devices, which can also be used to integrate the optics with electronics~\cite{Sibson2017_NC}. The current challenge in chip-scale QKD is to integrate the detectors onto the chips alongside the other optical components. Chip scale transmitters are not as bright as bulk optic sources and chip scale receiver systems will often need single mode fibre/waveguide coupling~\cite{Canning:19}. Chip scale devices also do not solve the requirements of auxiliary systems such as fine pointing and beam expansion. However, longer term road-maps predict the proliferation of tiny (few gram) satellites that can be launched and propelled using ground or space-based lasers~\cite{millan2019small} so such platforms may be cheap and plentiful for future quantum networks if these challenges could be solved.


\subsection{Environment, radiation, thermal, vacuum, shock/vibration.}
\label{subsec:enviromental_factors}

\noindent
Single photons detectors are susceptible to damage induced by space radiation~\cite{anisimova2017mitigating} and require to be cooled to below -30$^\circ$ C to operate. Radiation hardened electronics and devices can not be immediately adopted for use in low-cost missions that use microsatellites. However, radiation damage can be minimised by a number of methods. The prominent method is to use radiation shielding, which increases the size and weight of the satellite. Also, by implementing tighter opto-mechanical design, it is possible to reduce the cross-sectional area of susceptible devices that are exposed to radiation. Alternative methods include active cooling of the detectors, annealing, and reducing the bias voltage (degraded detector performance)~\cite{tan2013silicon}. Cooling of the detectors can be achieved by passive radiators which may reduce the area available for solar panels.

Entanglement sources often rely on temperature sensitive, 3 or 4 wave mixing processes in non-linear optical media. This necessitates temperature stabilisation that contributes to the total SWaP. For typical Spontaneous Parametric Down Conversion (SPDC) based sources the typical wavelength change is about 0.3\,nm$/^\circ$C~\cite{joshi2014entangled}. The wavelength change of the pump laser will also contribute to a shift in the signal and idler wavelengths.

\begin{figure}[t!]
\centering
\includegraphics[width =0.98\columnwidth]{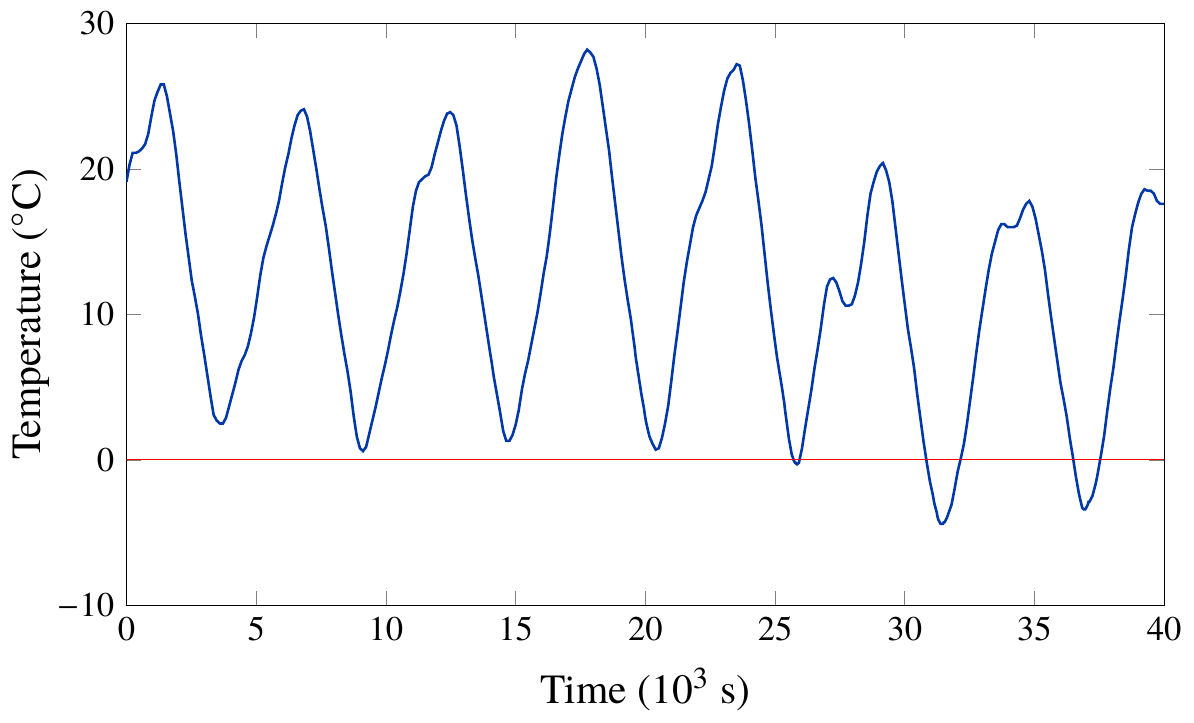}
\caption{In-orbit temperature fluctuations of SpooQy-1~\cite{Villar2020_O}. This 3U CubeSat was deployed from the ISS (400 km altitude) in 2019 and orbited the Earth every 90 min (5400 s). The temperature is recorded by the on-board computer for time period UTC2019-12-26 06:39:00 to UTC2019-12-26 20:12:30. Temperature fluctuation were measured  within the range of -10$^\circ$C to +40$^\circ$C. The average temperature depends on the fraction of the orbit it is in the Earth's shadow as well as the amount of internal power dissipation.}
\label{fig:thermal_fluctuations}
\end{figure}

Additionally, the design of optical systems must account for other space conditions. For example, the surface of a satellite can exhibit thermal fluctuations that have an extreme range such as -170$^\circ$C to 120$^\circ$C depending on orientation and time spent in the sun~\cite{Villar2020_O}. However, for a satellite in LEO, the temperature extremes are significantly smaller as illustrated in Fig.~\ref{fig:thermal_fluctuations}. The design of microsatellites must account for this fluctuation to accommodate thermal expansions that can also affect sensitive optical alignments. Optical elements should also be robust to vibrations during launch and operation.

For the most part, quantum communication has proven to be difficult in daylight conditions because bright background light can cause significant noise on the detectors. Thus satellite quantum communications are currently only possible during night from an OGS location with minimal light pollution to a satellite in eclipse. The necessity for dark skies limits the availability of suitable OGS locations that would allow the best visibility of the satellite, contribute minimal noise, and which are relatively close to population centers~\cite{Mazzarella2020_C}.

%

\subsection{Optical link configuration}
\label{subsec:optical_link_considerations}

\noindent
Implementing a QKD link requires a choice for the photon encoding, operating wavelength, photon bandwidth, and detectors. Mission budget aside, the optimal choice of these optical and system parameters have a strong dependency on the environment of the optical link. Therefore, a general statement on the optimal configuration of the link is not possible. Instead, we review the trade-offs to consider in configuring each parameter.

First, there are many signal encodings that can be chosen. For DV protocols, signals can be encoded into different photon degrees of freedom. These include polarisation, time-bin, frequency, orbital angular momentum (OAM), or spatial modes of qubits or qudits. For applications in long-distance free-space and satellite QKD, OAM and spatial mode encodings are not suitable since they are susceptible to atmospheric turbulence~\cite{krenn2016pnas}. Notably, OAM suffers from large divergence at long-distance propagations~\cite{padgett2015njp,vallone2016pra}. Instead, polarisation or time-bin (or frequency) encodings are a natural choice given their robustness to atmospheric losses~\cite{Vallone2015_PRL,Liao2017_N,Liao2018_PRL,vallone2016,jeongwan2018pra,jin2019oe}. For other applications, these encodings have differing performances that may impact the attainable key rates. For example, with polarisation being described by a bidimensional space, it is unsuitable for qudit encoding. Instead, time-bin, frequency, and OAM spatial mode encodings are better suited. Similarly, the use of time-bin encoding for satellite links has a drawback of requiring a compensation for the Doppler effect due to the Satellite motion~\cite{vallone2016}. Further, for time-bin encoding with qudits, it is difficult to perform a true mutually unbiased measurement.  This can however be circumvented by adopting a hybrid DV-CV approach and using a continuous variable as a second basis~\cite{Zhong2015}.

Second, the operating wavelength should be chosen to minimise different mitigating factors such as diffraction losses, turbulence sensitivity that differs with varying link configuration (i.e. uplink or downlink), scattering, absorption, and background light levels. Longer wavelengths have the advantage of reduced background photons from light pollution or sunlight scattering during day time operation but suffer worse diffraction losses than shorter wavelengths. There are several atmospheric transparency windows in the near infra-red and short-wave (telecommunication) infra-red regions that are usually considered due to the availability and performance of sources, detectors, and optical components operating at these wavelengths~\cite{kaushal2015}. The performance of the various QKD components at the chosen wavelength must also be considered. Specifically, the main variation in the attainable key rates with the optical wavelength is detector performance to noise signals induced by dark counts~\cite{Bourgoin:2013fk}. Visible-near infrared (Vis/NIR) silicon detectors have lower noise levels than InGaAs sensors operating in the short-wave infrared (SWIR) region. Superconducting nanowire single photon detectors can offer high detection efficiencies and low dark count rates over a wide range of wavelengths (Vis-SWIR) with the trade-off of greater cost, size, and complexity.

Finally, quantum communication satellites may need to perform inter-satellite communication. In such scenarios, the complications due to the relative motion of satellites (including accurate time synchronisation, pointing \& tracking, and link availability) are magnified. For untrusted node operation, satellites must be capable of maintaining at least two active optical links simultaneously~\cite{Yin2020_N}. Such satellite platforms are suitable for distributing entanglement, acting as quantum repeaters, performing measurement device independent protocols and quantum teleportation. For smaller satellites, coarse pointing of a single telescope on CubeSats is achieved by body pointing through reaction wheels. However, the disturbance caused by pointing a second telescope poses a major challenge to the design of smaller satellites with dual optical links.


\subsection{Timing and synchronisation}
\label{subsec:timing_&_sync}

\noindent
Correlating the transmission and reception of quantum signals is important for satellite QKD. Long distances, relative motion, and uncertainties in time of arrival are significant challenges. The high loss typical of such links complicates matters. Dispersion is not an issue for propagation through vacuum, though there can be minor effects from the atmosphere in space-ground links~\cite{jian2012real}.

Most quantum communication protocols account for losses by having the receiver(s) report arrival times of detected photons. Users then compute the temporal-cross-correlation to identify appropriate detection events. To avoid errors, they must precisely record the arrival time of all photons with a high relative accuracy. Timing jitter in the detectors and transmitters typically limits this relative accuracy to about 50 or 100 ps, though detector jitter of under 3 ps has been reported in a SNSPD~\cite{korzh2020demonstration} and 1 ps is demonstrated possible using optical gating~\cite{bouchard2021achieving}. The transmitter(s) and receiver(s) are also limited by the precision of their individual local clocks and the clock signal they share for synchronisation purposes. A common strategy is to encode the time synchronisation information onto either the uplink or downlink beacon lasers and discipline all local clocks to this signal~\cite{Zhang2020_conference}. 

While the beacon laser allows users to share a frequency reference, a small percentage of detection events must be utilised to continually compensate for the varying path length between the OGS and the satellite. This is because of the rapidly moving satellite (at roughly 1$^\circ$/s or linear speeds of up to $\approx$ 8 km/s)~\cite{agnesi2019sub}. Relativistic effects and other clock drifts can also be compensated in this manner. Those detection events used for compensating effective path length differences should not be re-used to generate the secure key to help limit the information that can possibly be leaked to an adversary.  

Typically, quantum communication requires some sort of shared reference frame to enable coherence to be observable~\cite{bartlett2007reference}. Beacons can be used to provide alignment information, e.g a linearly polarised laser allows the receiver to compensate for any relative rotation around the line of sight direction. If the change in alignment is slow, reference frame independent protocols can be used to eliminate the need for such alignment steps~\cite{laing2010reference,tannous2019demonstration}. In CV QKD protocols, the reference frame (local oscillator phase) can be transferred in the form of a strong reference pulse that is transmitted along with the quantum signal (e.g. using orthogonal polarisations) that can then be mixed to perform a coherent (homodyne~\cite{PhysRevLett.88.057902}, intradyne~\cite{laudenbach2019pilot}, or heterodyne~\cite{PhysRevLett.93.170504,Costa2006_SPIE}) measurement by the receiver. Alternative methods of transferring this phase reference are also possible~\cite{laudenbach2019pilot} but depending on the quantum degree of freedom (e.g. polarisation, phase, time-bin) being used to encode information, a method of reconciling reference frames is required.


\subsection{Classical communications and data processing}
\label{subsec:data_transfer_processing}

\noindent
Quantum communication protocols often require supplementary classical communication channels. For example, QKD requires authenticated public channel for reconciliation and other post-processing tasks to ensure the security of messages. The overhead can be significant, requiring high-speed and low latency classical communication. For correlation of transmitted and detected signals, time stamp data that has to be transferred can be generated at several megabytes per second for high-speed sources and loss-loss channels. For practical QKD systems, reconciliation should occur in real-time to minimise latency in key generation, hence the need for high-speed classical communications channels that can operate simultaneously during an over pass. Error correction and privacy amplification also generates extra traffic. For small satellites, this can be challenging though the commercial availability of X-band and even laser communications sub-systems for CubeSats may alleviate this bottleneck~\cite{NASASOTASST}.

Quantum communication protocols may also require significant data processing for tasks such as signal extraction, error correction, and privacy amplification. Computational power and memory are often constrained on spacecraft thus it is preferable to offload as much processing tasks to the ground segment. These limitations are particularly acute in small satellites~\cite{salas2014phonesat}.


\section{Improving space-based quantum technologies}
\label{sec:enabling_tech}

\noindent
In this section, we review improvements to key enabling technologies that could improve the performance of satellite-based quantum communications. We also provide a perspective on system-level changes that could deliver improvements independent of developments to individual components.


\subsection{Optical systems}
\label{subsec:optical_systems}

\noindent
Satellite quantum communications necessarily requires long-distance free-space optical links. Losses in the optical channel are reduced by increasing transmission aperture diameter for narrow beam divergence, increasing receiver aperture diameter to increase collection area, and improving beam pointing to maximise received intensity.

A free-space optical link uses telescopes to both transmit and receive quantum signals. Depending on the configuration of the communication system (see Fig.~\ref{fig:satellite_gs_links}), the telescope may be situation in space or on the ground. Space-based telescopes need to be rugged, compact and easily integrated with a different quantum hardware. New materials, such as Silicon Carbide, enable lightweight, rigid, and thermally stable mirror substrates, that can reduce the total satellite mass and enable more optically ambitious designs~\cite{Fruit2017_SPIE}. One option for increasing the performance of space-based telescopes within a limited SWaP budget is to use deployable optical elements that are stowed on launch and moved into place once the satellite is in orbit~\cite{Corbacho2020}. For example, the work of~\cite{Schwartz2016} increases the telescope area by four times (increasing the amount of light transmitted/collected) and the baseline by three times (decreasing the amount of diffraction) in 1.5U of CubeSat volume.

Ground-based telescopes are usually re-purposed astronomical telescopes or satellite laser ranging observatories, which perform well for communications purposes but are quite expensive~\cite{stepp2003estimating}. Efforts have been made to redesign satellite quantum communication OGSs from the ground up without the need for astronomical observation~\cite{moll2015ground}. These unnecessary requirements include wide-field diffraction-limited imaging performance, sub-arcsecond level sidereal tracking by the telescope mount, and broadband spectral performance that are superfluous for satellite quantum communications. Mass production of ground optical terminals should also lead to cost reductions and to considerable saving on total system costs~\cite{innovateUK2018}.

Due to the narrow beam divergence of the transmitted quantum optical signal, high precision pointing is required to successfully convey weak optical signals across long distances. This is achieved by the acquisition, pointing, and tracking (APT) system that usually incorporates fast-steering mirrors (FSM), beacons, and tracking sensors to provide the microradian-level accuracy needed (see Section~\ref{subsec:loss_sources}). These systems increase SWaP demands and are critical failure items. Alternative APT components have seen continual advancement in miniaturisation, price, and performance, leading to greater accessibility to smaller missions~\cite{Shinshi2020_E,Knapp2020_AJ}. Attitude determination and control systems (ADCS) for small satellites, usually employed to perform initial coarse pointing, have seen rapid improvement in performance levels. This has reached the level whereby beaconless (open-loop) body-only pointing can enable high-speed laser communications~\cite{rose2019optical} without the need of a fast steering mirror. If further improvements are possible, this increases the scope for reduction in cost and system complexity of quantum communications.

As an alternative to mechanically steered beams, solid-state solutions are under development~\cite{ESAITT} that eliminate moving parts by the use of electro-optical modulators~\cite{Wu2019_NC} or through phased-array beam-forming techniques~\cite{Gozzard2020_OL}. There is the potential from spin-in from automotive LiDAR development to further reduce costs and improve miniaturisation~\cite{Miller2020_O}. These may allow simplification of mechanical design, increased reliability, and the development of novel optical configurations that have operational or performance advantages over conventional layouts.

The addition of adaptive optics (AO) may improve the link quality and resilience of the optical channel to turbulence and background light. This is mainly of use in a downlink configuration (Fig.~\ref{fig:satellite_gs_links}b) where wavefronts arriving at the OGS are distorted by refractive index variations. These distortions cause distorted focused spots that can lead to increased losses through imperfect coupling to detectors. For some encodings, e.g. time-bin, turbulence-induced wavefront errors can also lead to quantum bit errors unless special receivers are used~\cite{jin2019oe}. Higher order wavefront correction (beyond tip-tilt) utilises a wavefront sensor and deformable mirror to undo the wavefront distortion and allow for restoration of a small point spread functions (PSF) and high Strehl ratio~\cite{weyrauch2002fiber}. This is especially important for coupling incoming quantum signals to small diameter detectors or single mode fibres and for allowing for aggressive spatial mode filtering to reduce background light~\cite{Hughes2002_IOP,gruneisen2020adaptive}.

Adaptive optics can also be considered in uplink configurations to pre-distort a transmitted beam wavefront based on prior knowledge of the turbulence from analysis of an incoming beacon. This distortion is then reversed as the beam propagates towards the satellite, which can reduce beam spreading, wander, and loss. However, the effectiveness of this method is limited by the isoplanatic patch size and timescales over which the turbulence occurs, compared with the propagation delay over satellite-Earth distances~\cite{Pugh2020_AOT}. Adaptive optics systems (apart from tip-tilt fine-pointing) is of relatively limited use on the satellite itself, since the wavefronts are effectively spherical after propagating from the edge of the atmosphere to the position of the spacecraft. However, Microelectromechanical (MEMS) systems-based deformable mirrors, such as in the CubeSat DeMi mission~\cite{Morgan2019} could be integrated into the optics system to provide active correction of optical aberrations in flight due to thermal distortion or changes caused from launch vibration. To fit with the SWaP constraints of a smaller satellites, these devices and optical systems would require significant miniaturisation before becoming practical.

Recent developments in conventional free-space optical communications (FSOC) for satellite applications will become key enablers for quantum space communications. Most FSOC sub-systems can be adapted with minor modifications, such as APT systems~\cite{podmore2019optical}. Other FSOC sub-systems include timing and synchronisation, and high data bandwidth FSOC channels required for auxiliary classical channels required for quantum communication protocols (see Sections~\ref{subsec:timing_&_sync} and~\ref{subsec:data_transfer_processing}). Optical terminals for the European data relay service (EDRS) have been tested for QKD compatibility with the aim of incorporating such functionality with only minor modification~\cite{elser2017quantum}. Conversely, the narrow beam widths and stringent pointing requirements of QKD naturally boost the classical communication performance. Thus, integration of QKD functionality into space FSOC systems may bring further benefits aside from the security of QKD.


\subsection{Classical communication systems}
\label{subsec:laser_systems}

\noindent
As just discussed above and in section~\ref{subsec:data_transfer_processing}, quantum communication protocols often require a large exchange of classical data. Given bandwidth requirements, this naturally drives the use of shorter wavelengths for satellite communications, with the progression of commercially available smallsat radios operating in the VHF, UHF, S-Band and now X-Band regions~\cite{Alessandra2020_Inf}.

Laser communication can offer significant improvements over radio frequency communication, due to a combination of smaller beam divergence for a given transmit aperture, hence lower free-space loss, as well as higher bandwidth of optical frequencies. This concept has been demonstrated in several missions. First, NASA's optical payload for lasercomm science (OPALS) mission uses a 2.5 W 1550 nm laser with a beam diameter of 2.2 cm, achieving a bandwidth of 50 Mbps~\cite{abrahamson2014}. Second, NICT's space optical communications research advanced technology satellite (SOCRATES) mission uses a 20 mW 1549nm laser with a beam diameter of 5 cm and a 175 mW, 976 nm laser with a beam diameter of 1 cm, achieving a bandwidth of 10 Mbps~\cite{carrascocasado2017}. Finally, NASA's planned laser communications relay demonstration (LCRD) mission, a 500 mW 1550 nm laser is used with a beam diameter of 10 cm, which is expected to exceed a bandwidth of 1 Gbps~\cite{nasalcrd}.


%
\begin{figure*}[t!]
\subfloat[Satellite trajectories.]{\includegraphics[width=0.42\linewidth]{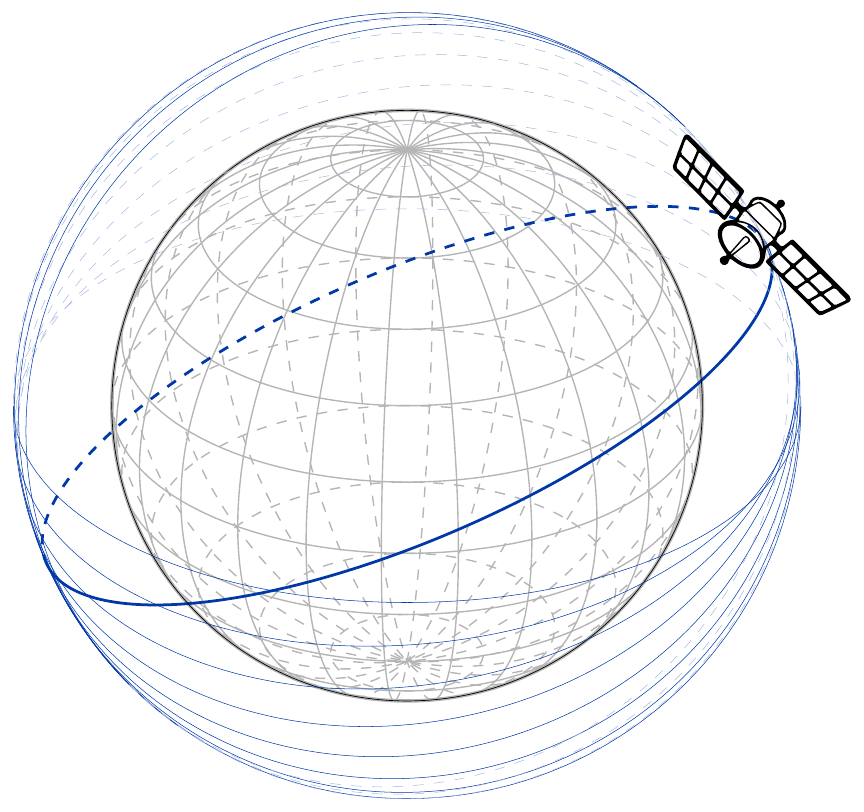}\label{fig:sat_trajectories}} \hspace{25pt}
\subfloat[General topology of OGSs and satellites.]{\includegraphics[width=0.53\linewidth]{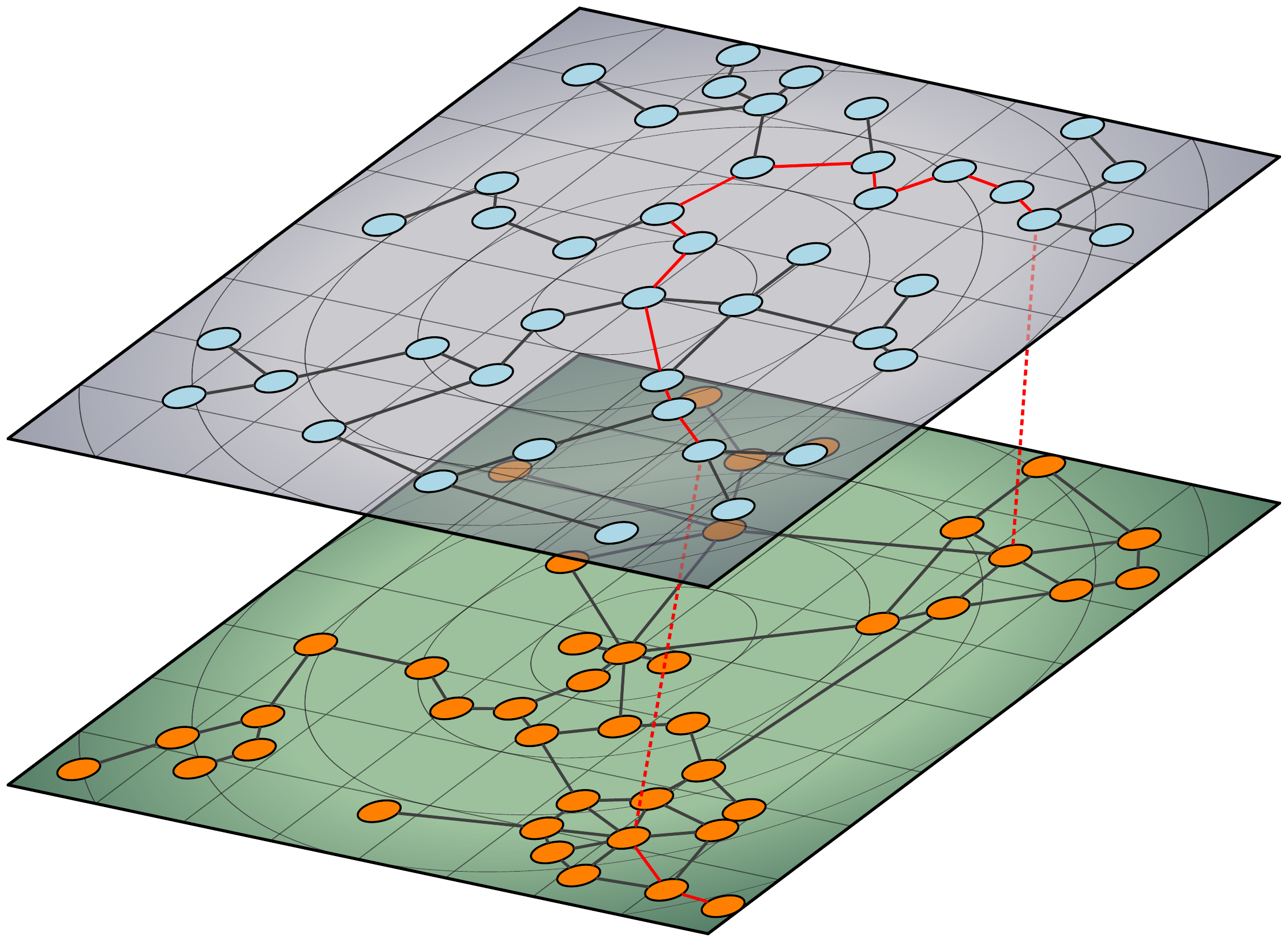} \label{fig:network}}
\caption{A global QKD network. Fig.~\ref{fig:sat_trajectories}: Satellites offer a way to extend the range of QKD to global distances by reducing photon absorption and distortion effects due to the Earth’s atmosphere. Through this approach, a single satellite can establish secure links between different OGSs separated by thousands of kilometers. A satellite can also act as a trusted courier for quantum keys. As illustrated here, a single satellite can pass over different OGSs, which permits the sharing of quantum keys with different OGSs. The payload computer transmits a bitwise sum of keys to the two respective ground stations using classical communications, and this enables the ground sites to generate a secure key between them, regardless of their distance. Fig.~\ref{fig:network} illustrates a general network of satellites and OGSs to realise global quantum communications. Each satellite and OGS act as a network node and are linked together to form repeater chains. This naturalises transmission links to arbitrary quantum networks over global scales that can be analysed using deep results from classical network theory. This can help choose optimised OGS ground network and satellite coverage to distribute entanglement in a resource efficient way. In such a network, OGS transmit quantum signals to the satellites, and through Bell measurement, each satellite can perform entanglement swapping and purification or measurement-device-independent QKD.}
\label{fig:Sat_QKD_network}
\end{figure*}%

\subsection{Sources}
\label{subsec:sources}

\noindent
Quantum communications typically require a source of quantum signal states to be transmitted. For prepare and measure trusted-node QKD~\cite{Vergoossen2020_AA}, the signals typically comprise single photon level states (for DV QKD) or coherent states (for CV QKD). For untrusted-node QKD, usually entangled photon pairs sources are used, though measurement device independent (MDI) QKD utilises non-entangled sources (see Ref.~\cite{QKDreview2020} for an overview of different protocols). For non-entangled DV protocols, true single photon sources are ideal, however these are at a relatively low level of technical maturity, hence lasers are used instead. The BB84~\cite{BB84} Weak Coherent Pulse (WCP) Decoy state (DSP)~\cite{PhysRevLett.91.057901,PhysRevLett.94.230503,PhysRevLett.94.230504,PhysRevA.72.012322}, and Coherent One-Way (COW) protocols~\cite{stucki2009continuous} are the most popular protocols to use phase-randomised WCP sources. Here, a laser pulse is attenuated such that the mean photon number per pulse is typically less than one to approximate an ideal single photon source. Due to Poissonian statistics, a WCP source has a non-zero multi-photon emission probability leading to information leakage to an eavesdropper but DSPs maintain security despite this~\cite{PhysRevLett.91.057901}. Historically, WCP sources are the simplest to implement and are a robust choice for satellites~\cite{Liao2018_PRL}. However, continuous variable and entanglement-based sources are reaching sufficient technological maturity for use in QKD applications~\cite{Silverstone2015_NC}. WCPs face some implementation flaws such as side-channel information leakage in auxiliary degrees of freedom, e.g. imperfect spectral or temporal overlap of pulses~\cite{nauerth2009information} that can compromise the security of the implementation. However, WCP sources have been used for the highest transmission rates so far, potentially up to GHz repetition in chip-based platforms~\cite{Sibson2017_NC}. Under development are true single photon sources, such as those based on quantum dots in 2D materials~\cite{palaciosberraquero2016, Vogl2019}. These sources do not require cryogenic temperatures and are also of interest to the wider quantum technology field. Specifically, highly bright on-demand sources would have many applications as can be used to improve quantum communication links.

Entanglement-based sources offer the advantage of verifiable security in untrusted-node QKD configurations (see Fig~\ref{fig:satellite_gs_links}c). Here the users can perform a Bell test to verify that the quantum states they share are in fact entangled and therefore the link is secure~\cite{Ekert1991_PRL}. In practice, the Bell test is often omitted and an entangled version of BB84 is implemented instead~\cite{Bennett1992_PRL}. Currently, a practical source for entangled photon pairs is through SPDC. However their brightness must be traded against quality, with the pump power being limited to a level below which excessive probability that multiple photons are emitted within a coincidence window degrades entanglement visibility~\cite{ecker2021strategies}. While the brightness of entangled sources is fundamentally limited by the damage threshold of the nonlinear optical crystals used, the practical limit in most cases is either due to saturation of single photon detectors (in unbalanced configurations e.g. local measurement of one half of each pair) or due to detection jitter that limits the coincidence detection accuracy. Waveguide or four wave mixing based sources~\cite{fulconis2005high} can also generate entangled photon pairs, but their rate is limited compared with lower order non-linear processes~\cite{Anwar2020_arxiv}.
To achieve a given key rate, an entanglement-based source needs to be much brighter than a typical WCP source, but entanglement protocols like E91~\cite{Ekert1991_PRL} can still produce keys under higher losses than WCP sources running DSP~\cite{neumann2018q}. Entangled sources with high quality and brightness would greatly improve to the quality of the quantum channel~\cite{Oi2017_EPJ}.

Using entanglement, higher dimensional, large alphabets and hyper entangled states can be used for dense coding~\cite{PhysRevLett.69.2881} i.e. to transmit more than one bit of information per photon or pulse. While quantum communication can be implemented with or without entanglement, several other quantum protocols rely on entanglement distribution, and this is a strong motivation to develop satellite-based entanglement distribution networks as illustrated in Fig.~\ref{fig:Sat_QKD_network}.

CV QKD protocols encode information in continuous degrees of freedom~\cite{PhysRevA.61.010303,PhysRevA.61.022309,PhysRevA.63.052311}, such as the position and momentum quadratures of the electromagnetic field~\cite{QKDreview2020}. A key difference between CV and DV protocols is that CV protocols use infinite-dimensional Hilbert spaces, so that they can encode more information per pulse and potentially achieve much higher key rates than DV protocols (e.g., reaching the PLOB bound~\cite{PLOB}). Typically, CV sources are based on Gaussian-modulated coherent states, which are easy to generate for the transmitter and are also easy to detect by the receiver (see Section~\ref{subsec:timing_&_sync}). Other schemes may involve the use of a middle (generally untrusted) relay performing a CV Bell detection, as is the case of the CV MDI protocol with coherent states~\cite{CV-MDI-QKD}. Currently, CV QKD sources and protocols are widely used for terrestrial links~\cite{Zhang19} achieving distances of the order of $200$~km in optical fibre~\cite{PhysRevLett.125.010502}. The security proofs of CV QKD protocols have recently been extended to the composable finite-size scenario, i.e. the most general level of security~\cite{QKDreview2020}.

Traditionally, quantum sources have been developed using bulk optics~\cite{bedington2015deploying,cao2019satellite}. For space applications, miniaturisation and robustness of integrated photonic circuits are attractive. There has been a general trend towards fabrication of sources as well as optics using using photonic chips~\cite{PhysRevX.8.021009,Sibson:17}. However, the control electronics and ancillary systems will also need miniaturisation in order to derive full benefit from this technology~\cite{Vest2015}.


\subsection{Detectors}
\label{subsec:detectors}

\noindent
For ground station receivers and DV systems, the current state-of-the-art in single photon detection provides sufficient detection efficiency (DE), dark count rates (DCR) and timing jitter. Detection efficiency increases general system throughput which directly increases key rates and helps overcome finite key effects in a shorter amount of time~\cite{Sidhu2020}. Dark counts contribute to the total detected error rate and should be minimised, while timing jitter contributes to an uncertainty in the detection event time and restricts the amount of temporal filtering that can be applied to reject spurious counts. This constrains the maximum repetition rate of the system so that each detection event can be distinguished.

Current QKD systems generally use either high performance single photon avalanche diodes (SPADs)~\cite{bronzi2016,ceccarelli2020recent} or higher performance and higher cost superconducting nanowire single photon detectors (SNSPDs)~\cite{esmaeilzadeh2017,You2020_N}. Generally, SNSPDs have lower jitter, can exhibit higher photon detection efficiencies (PDE) especially at longer wavelengths (e.g. 1.55 micron), but with the trade-off of higher costs, SWaP, and lower ease of deployment. SNSPDs require cryogenic cooling for their operation, though SPADs benefit from thermoelectric cooling to reduce dark count. The development of arrays of single photon detectors opens up new possibilities such as photon number resolution~\cite{jiang2007photon} and wide-field of view receivers~\cite{Donaldson_2021}. SPAD arrays are reaching a high level of maturity, in part due to demand in autonomous vehicle LiDAR~\cite{matsubara2012development}, though SNSPD arrays are also under development~\cite{wollman2019kilopixel}. 

Space-based detectors are required for different configurations such as uplink, trusted-node entanglement downlink, and any inter-satellite quantum communications. Besides the need for SWaP improvements for commercial devices~\cite{tan13}, there is the additional need for radiation tolerance and dark count suppression through shielding~\cite{yang2019spaceborne} and laser annealing~\cite{lim2017laser}.

Currently available components are sufficient for CV ground station receivers~\cite{gunthner17}, although the homodyne detectors used in CV QKD are not at the same level of commercial maturity as detectors for DV systems. For CV detectors, homodyne-based detection relaxes the performance requirements of the photodetectors~\cite{PhysRevLett.88.057902,Costa2006_SPIE,laudenbach2018continuous}. 


\subsection{Quantum memories}
\label{subsec:memories}

\noindent
Quantum memories (QMs)~\cite{Lvovsky2009, Zhao2009, Gundogan2015} are of central importance to many protocols in quantum information processing where synchronisation of otherwise probabilistic events are needed. One application on this is in quantum communications, where QMs could act as nodes of quantum repeaters. However, practical quantum repeaters are still far from being realised in the laboratory due to their technically demanding nature. Therefore, efforts have been largely focused on implementing memory-assisted (MA) QKD~\cite{Abruzzo2014, Panayi2014, Luong2016} protocols, which are the simplest use-case of QMs in quantum communications. The main idea is to divide the total communication distance into two segments separated by a central station housing two QMs. For an inward scheme, Alice and Bob prepare single photons as per the usual BB84 setting and send them towards a central station where they are stored in QMs~\cite{Abruzzo2014, Panayi2014}. A Bell-state measurement (BSM) is then performed on the memories to extract a secret key.  Conversely, for an outward scheme, single photons entangled with the internal atomic states of the QMs at the central station are sent to the communicating parties~\cite{Luong2016}. A BSM is then performed, again upon the successful detection of a single photon by Alice and Bob to extract a secret key. The use of QMs in these protocols allow the key rate to scale with $\sqrt{\eta_\text{ch}}$, therefore outperforming the limit of $1.44 \eta_\text{ch}$ bits per use for a direct communication link~\cite{PLOB}. Ideally, one could reach the single-repeater bound $-\log_{2}(1-\sqrt{\eta_\text{ch}})$~\cite{Pirandola2019_CP}. The first experimental MA-QKD work has only recently been performed with an SiV centre in diamond cooled down to $\sim$100 mK as a QM~\cite{Bhaskar2019}. 
These ideas can be extended to space-based scenarios to increase the attainable key rates in a line of sight setting. Storage time of around $\sim$100 ms in combination with $>50\%$ storage efficiency would increase the key rate by an order of magnitude~\cite{Gundogan2020_arxiv} over direct entanglement distribution protocols~\cite{Ekert1991_PRL, Ma2007} with an inwards (uplink) scheme. The outward (downlink) scheme experiences less loss; however, it requires significantly longer storage times, in the order of seconds, together with a temporal multiplexing capability to store up to 1000 temporal modes~\cite{Gundogan2020_arxiv}. The required storage time can be reduced further with many memory pairs operating in parallel, as proposed in Ref.~\cite{Trenyi2020}. Unlike the uplink scheme, the downlink protocol could be extended into a full space-based repeater architecture with further entanglement swapping operations between the neighbouring nodes. 

As discussed in Sec.~\ref{subsec:entanglement_networks}, a full quantum repeater network realized with the help of QMs is essential for extending the range of untrusted quantum networks to truly global ($>10^4$ km) distances~\cite{Boone2015, Gundogan2020_arxiv, Liorni2020_arxiv}. In this context, quantum memories on board satellites could bring several advantages over hybrid schemes where the memories are located in ground stations. First, it reduces the number of space to ground links from $2^{(n+1)}$ where $n$ is the nesting level to 2. Likewise, the number of ground stations is reduced from $2^{(n)+1}$ to 2. This greatly relaxes the requirement that all ground stations should simultaneously have good weather conditions. Second, space-ground links may suffer from the atmospheric loss which becomes dominant at small grazing angles as the channel length increases. Space-space links, on the other hand, are mainly limited by the beam diffraction and available receiver aperture. Lastly, space-space links would suffer significantly less from Doppler shifts which should be compensated to ensure the indistinguishability of the photons for the BSM.  In light of these advantages, recent analyses have demonstrated that utilizing QMs on board would facilitate entanglement distribution across global distances in time a scale of $\sim$1~s~\cite{Gundogan2020_arxiv, Liorni2020_arxiv}, which is at least 4 orders of magnitude faster than hybrid ground-space repeater links.

Currently, the technically demanding nature of the QM experiments have stimulated research towards a hybrid satellite-ground QR network architectures \cite{Boone2015,Sumeet2019}, where QMs are located in ground stations. However, promising recent technical advances in the field warrant attention to the ideas presented in this section. These advances take place in three independent fronts. First, the rapidly developing field of space-based atomic physics experiments for gravity sensing, atom interferometry and optical clocks. These efforts have culminated in the realisation of the first Bose-Einstein condensate (BEC) in space~\cite{Becker2018} in a suborbital flight and operation of a BEC experiment on-board the ISS~\cite{Aveline2020}. Second, significant advances in the performances of QMs have been demonstrated. Sub-second long storage times have been successfully combined with high efficiencies in the single-photon regime~\cite{Yang2016} whereas classical light has been stored for around a second in a warm vapour~\cite{Katz2018}, for up to a minute in a cold-atom based memory~\cite{Dudin2013} and very recently for up to an hour in a rare-earth ion doped (REID) crystal~\cite{Ma2021}. Finally, integration and miniaturisation of these devices is another direction that many researchers are pursuing. These efforts include on-chip memories based on single colour centres in diamond~\cite{Wan2020}, laser written waveguide QMs in REID crystals~\cite{Seri2019} and atom-chip quantum memories~\cite{Keil2016}. Thus one can hope that operation of such memory devices in space will soon be within reach.

\subsubsection{Frequency conversion}
\label{subsubsec:FreqConv}

\noindent
Coherent frequency conversion of optical signals is desirable for integrating space and terrestrial networks as current hybrid networks rely on measurement and retransmission~\cite{Chen2021_N}. Quantum coupling of free-space and fibre links would allow for untrusted OGSs as access points between space and ground, and entanglement distribution via satellites and optical fibres~\cite{gruneisen2017modeling}. Entanglement distribution could be via the use of quantum repeaters equipped with memories in each OGS to relay the quantum states to other users in the network (or satellite) for a downlink (or uplink) scenario.

The wavelengths (frequencies) employed for the satellite link may differ from the ideal wavelengths for either onward long-distance fibre-based QKD or for coupling into quantum memories that require atomically narrow transitions~\cite{Radnaev2010}. Therefore, coherent frequency conversion will become a very important technology to match the requirements of the different network segments. Multimode wavelength conversion is possible but single-mode operation may be dictated by downstream utilisation, coupling into single mode fibres for example~\cite{gruneisen2017modeling}.

1991So far, frequency conversion has been mainly demonstrated with nonlinear crystals either in bulk or waveguide geometry~\cite{allgaier2017highly,dreau2018quantum, bock2018high, ates2012two,zaske2012visible, Maring2017}. Alternative systems such as atomic ensembles have also been used for conversion from quantum memory wavelengths to telecommunication bands~\cite{Radnaev2010}. Internal conversion efficiencies of these experiments were limited to around 50\%, however this limit has recently been pushed to an unprecedented 96\% in an experiment where long-distance entanglement distribution has been demonstrated between Rb atoms and telecommunication photons at 1522 nm~\cite{vanLeent2020}. Upconversion can also be used to allow for the detection of 1550nm signals using silicon detectors by first transforming the longer wavelength photons into shorter wavelength photons that are within the absorption band of cheaper Si-SPADs~\cite{Liao2017}.

\subsubsection{Vacuum}
\label{subsubsec:vacuum}

Many quantum platforms require ultra-high vacuum (UHV) conditions in order to operate, such as cold atomic gases and ions. Despite being in the vacuum of space, dedicated UHV systems are usually required to provide these conditions. There has not been a high demand for satellite UHV systems outside of quantum applications, hence the maturity of space UHV technology is relatively low.

The challenge for their development is mainly SWaP but also long term operation and reliability. In addition, the operation of the UHV system must not affect the rest of the payload, e.g. vibration from pumps or magnetics fields. There are efforts to produce small, robust, and low maintenance systems that are space and microgravity capable~\cite{Ren2015_V, elliott2018nasa, vogt2020evaporative}. Some approaches utilise chambers with passive getters~\cite{Abraham2016} whereas the cold atom laboratory on board ISS uses commercial components~\cite{elliott2018nasa,Aveline2020}. A custom made vacuum chamber with three layers of shielding has been utilised to demonstrate the first in-orbit operation of an optical atomic clock~\cite{Liu2018}. 
Alternatively, one could exploit the vacuum of space. This concept has been proposed for the macroscopic quantum resonators (MAQRO) project to test large-scale quantum superpositions at Lagrange point 2 which has conditions of deep space~\cite{kaltenbaek2020tests}.

In LEO, the particle density is typically too high for quantum cold gas experiments, but the wake shield effect has been proposed to generate UHV conditions external to a satellite~\cite{wuenscher1970unique,melfi1976molecular}. The high orbital velocity of a LEO satellite sweeps up gas particles, leaving an extremely low density region in its wake. A specially designed spacecraft could produce large UHV volumes with the added benefit of near infinite pumping speed due to the open nature of the chamber. This was tested by the Wake Shield Facility (WSF) launched by the Space Shuttle, though the results were inconclusive~\cite{strozier2001wake}.

\subsection{Cryogenic systems}
\label{subsec:cryogenic_systems}

\noindent
Low temperature operation is vital for devices such as superconducting detectors, solid state quantum memories, and quantum dot photon sources. This usually requires active cooling, though for some missions passive radiative cooling has been considered to avoid potential issues of pump vibration~\cite{Zanoni2016_ATE}. Cryogenic systems have been successfully deployed on missions such as Gravity Probe B~\cite{Everitt2015}, Planck~\cite{Morgante2009}, and Herschel~\cite{Collaudin2010}. The duration of these missions were limited by Helium boil-off and venting. To simplify deployment and operation, it is desirable to use closed-cycle coolers. Current developments include passive cooling~\cite{Hechenblaikner2014} that can reach down to 16~K by carefully designing the payload and isolating the experiment from any heat sources; active coolers for SNSPDs can cool detectors down to 2-4~K~\cite{You2018, Dang2019, Gemmell2017}. In addition to these efforts miniaturised cryocoolers are being developed for tactical and space applications~\cite{Olson2014} which with further development could be suitable for use on board nanosatellites.


\subsection{Clock Synchronisation}
\label{subsec:Clock}

\noindent
In quantum communication applications, precise synchronisation between Alice an Bob is required to correctly associate the symbols transmitted and also to reduce the quantum bit error rate (QBER). A sub-ns synchronisation between two terminals is required to fully exploit the high timing accuracy of current single photon detectors, which can reach few tens of ps. This can be achieved by sending periodic reference pulses between the two terminals~\cite{Liao2017_N}, however this solution requires the transmission of an additional signal thus increasing the complexity of the scheme. Alternative synchronisation protocols, which do not require any additional signal other than the quantum one, have been recently proposed~\cite{calderaro2019fast,lee2019symmetrical}.  These protocols exploit the quantum states to perform clock recovery and retrieve the beginning of the quantum sate transmission thus avoiding any other synchronisation signal. 


\subsection{Future prospects}
\label{subsec:future_prospects}

\noindent
In addition to improving individual components on board satellites, there are also system-wide implementations that could improve the performance of satellite QKD. In this section, we give an overview of some of these more speculative improvements that may deliver improvements to global quantum communications. 

First, multiple, independently steerable telescopes are necessary to distribute entanglement to multiple OGSs. This is essential for a quantum network with untrusted nodes, and in a trusted satellite network, to minimise latency for key generation when multiples ground stations are in view. Transmitting signals to multiple OGSs independently is also an important requirement for MDI~\cite{MDI-1,MDI-2,CV-MDI-QKD}, twin-field QKD~\cite{TF-QKD} and implementations that use quantum repeater-based and quantum memories. However, steering multiple telescopes on small satellites is difficult given the mechanical complexity and mass of such a system. Steering separate telescopes could affect alignment of other satellite and optical systems. These disturbances could be mitigated with twin tethered nanosatellites where control of one body-mounted telescope does not impart a momentum change to the other system.

Second, formation flying of small satellite clusters would extend the range of applications that could benefit from satellite systems. This would be particularly important for distributed applications such as quantum enhanced sensing. This would require on-board implementation of high precision inter satellite positioning, timing, and synchronisation, for both relative motional knowledge and control.

Finally, compensation of Doppler shifts due to rapid relative motion of satellites may be required for some applications. For example, the typical speed of a LEO satellite is 7800 ms$^{-1}$, which has a fractional Doppler shift of $\beta = v/c =  2.6\times10^{-5}$. Compensating for this shift is particularly important for systems using quantum memories, since the signal must couple with a narrow line width of the quantum memory. This may require active compensation and tracking of the shift in conjunction with inter-conversion between flying and static quantum systems. Generally, other optical systems required on a satellite include mechanisms that can efficiently couple free-space photons with single modes in QMs, but also for other applications such as BSMs for MDI QKD.


\section{Fundamental physics experiments}
\label{sec:fundamental_expts}

\noindent
Our knowledge of fundamental physics is paved by two theoretical frameworks. First, general relativity (GR) provides the most accurate description of gravity to date as a description of the geometry of space-time. Second, quantum mechanics (QM) provides a precise explanation for the physical properties of nature at the scales of atoms and subatomic particles.
Both have demonstrated resounding successes through experimentation verification. Despite this, a framework that treats both GR and QM consistently at the same time and provides a good understanding of fundamental physics across all scales has not been found . 
While a valid quantum theory of gravity could be a solution to this problem, any unified theory which aims to provide a consistent description of the universe would need to address the incompatibilities between the two. 
Progress towards a more fundamental understanding of physics requires experimental access to scales where quantum and general relativistic effects interplay. The advent of satellite-based quantum communications gives increased access to space, which enables experiments at larger distances, higher speeds, and with non-stationary detectors.
This provides a more flexible environment to test the interface of quantum and gravitational theories. Specifically, quantum theoretic predictions can be tested in curved spacetimes. 
Any deviations between these predictions and their experimental outcomes could indicate a route to unifying quantum theory with general relativity.

Satellites operating in LEO are currently realisable. These orbits permit access to distances greater than $10^6$ m and relative detector speeds of around $\sim$ $10^{-5} c$. Compared with terrestrial experiments, the longer free-fall times enable high-precision tests of GR and the equivalence principle for quantum systems. Space-based sources of entangled photons promise long-distance tests of quantum theory, quantum field theory in curved spacetimes, and the interplay between relativity and quantum entanglement. The list of possible experiments of fundamental physics increases with the accessible distance. Future missions of satellite-based quantum communications will enable experiments at the scale of interplanetary distances.

In this sections, we review the fundamental physics experiments that can be performed with access to a network of satellite-based quantum communications architecture. This includes performing tests of general relativity, quantum theory, and finding signatures of Physics beyond the standard model such as dark matter, modified theories of gravity and quantum field theory (QFT) in curved spacetimes. We categorise these experiments in terms of the {\it main feature} of satellite-based quantum networks they aim to investigate.  These are gravitational and relativistic effects, tests of the foundations of quantum mechanics and experiments that use satellite arrays with significant overlaps between them.  

\begin{figure}[t!]
\centering
\includegraphics[width =\columnwidth]{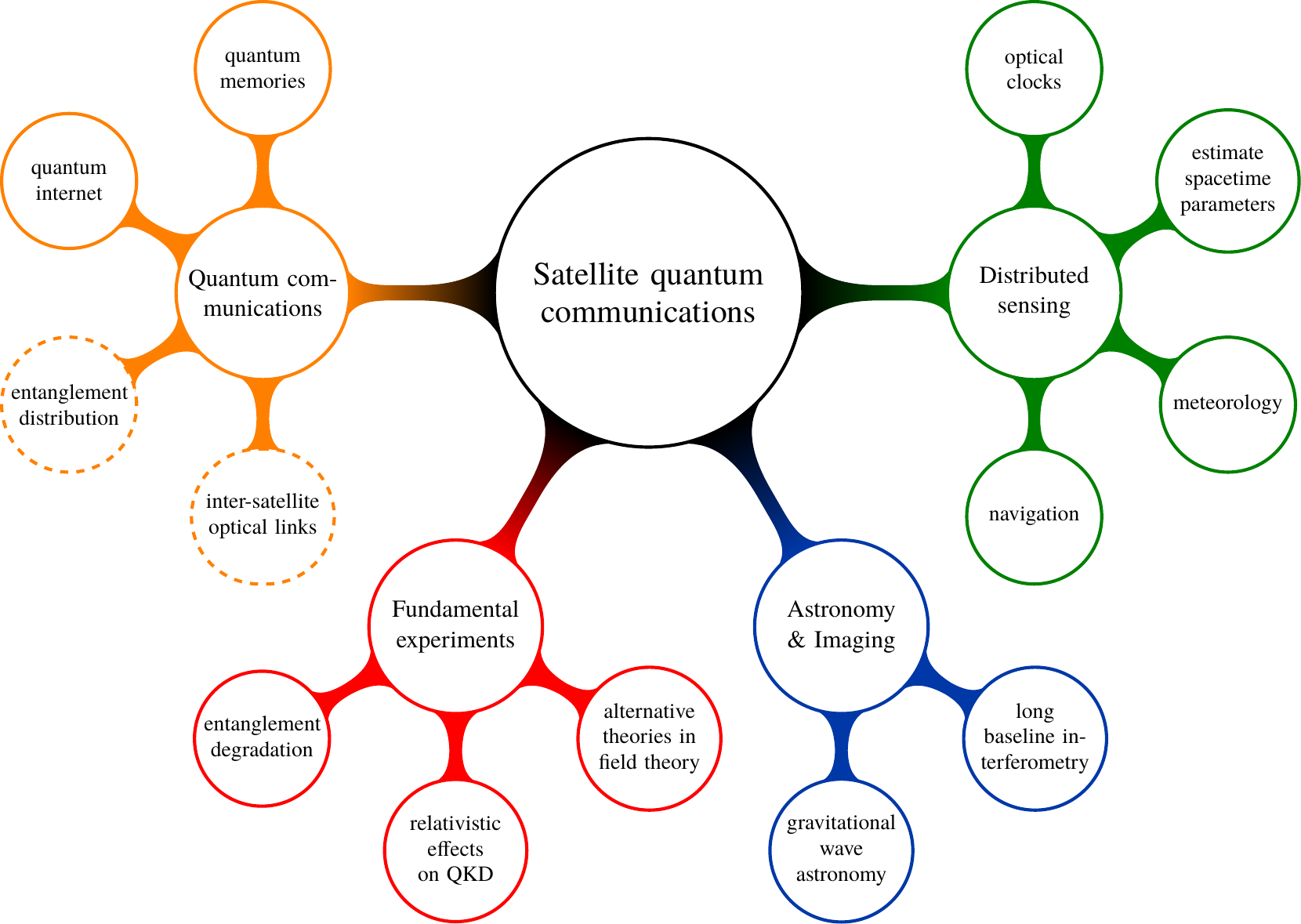}
\caption{Mindmap of future experiments in fundamental Physics that could take advantage of a satellite-based quantum communication platform. Different applications are colour-coded according to different quantum technologies. Dotted circles indicate proof-of-concept experiments that would enable further progress of technologies.}
\label{fig:expts_mindmap}
\end{figure}%

\subsection{General relativity and quantum field theory in curved spacetime}

Gravitational time dilation (or gravitational redshift) is one of the major predictions of GR. A consequence of this effect is that a clock on Earth will run at a slower rate relative to one in orbit on a satellite. Any deviation from this could provide hints towards a modification of GR.
While any discrepancy would be small, the development of satellite-based atomic clocks allow us to experimental test this effect.  
For example, in 1976, the Gravity Probe-A \cite{Quattrini2014} measured the frequency shift of a satellite-based atomic clock, relative to one on Earth.  The predictions of GR were verified with an uncertainty of $1.4\times 10^{-4}$. This agreement has been improved further by new experiments, such as those performed using the Galileo 6 and 7  satellites \cite{Hermann2018PRL, Delva2018PRL}, which decreased the uncertainty by a factor of 5.6.

A different approach to testing gravitational time dilation is to use satellite-based interferometry.  A Mach-Zehnder interferometer allows a photon to travel two different paths.  If the two paths correspond to different gravitational equipotentials, then gravitational time dilation results in a phase difference in the two paths, which is observed in the interference at the output.  
Similar effects are observed in matter interferometry, for instance the Collela, Overhauser and Werner (COW) experiment tested gravitational effects in matter interferometry within the Newtonian gravity regime \cite{colella1975}. Satellites allow us to increase the size of the interferometer, which improves the sensitivity such that we can detect effects beyond the Newtonian regime. A few variants of these experiments have been proposed, where a suitable degree of freedom serves as a local clock.  For instance, one could use  the position of the photon in the interferometer \cite{Zych2012} or the precession of polarisation  \cite{PhysRevD.91.064041}.  A practical implementation has been proposed using a``folded" interferometer with a single Earth orbiter and a ground station \cite{Pallister2017}. 

Satellite-based tests of the equivalence principle provide another method of investigating modifications of GR. In 2017 the MICROSCOPE \cite{Touboul2017PRL} mission used a microsatellite to investigate violations of the equivalence principle. 
The experiment verified the equivalence principle with a precision of order $10^{-15}$, 100 times better than that achieved using Earth-based experiments.

Rather than looking for discrepancies in GR, one could instead look directly at how quantum phenomena are affected by gravity.  The aim of Quantum Field Theory (QFT) in curved spacetime is to explore physics at the intersection of GR and QM, for relatively low energies regimes.  The resulting models give rise to effects such as Hawking Radiation or particle creation by an expanding universe. 
Satellites provide a powerful experimental platform to investigate the physics of QFT in curved spacetime.  For example, one can look at how curved spacetime affects the propagation of photons from satellites to Earth \cite{Bruschi2014_PRD,Kohlrus2017_EPJ}.  Similarly, photons entangled between a satellite and Earth can be used to study how gravitational differences, or relative  velocities / accelerations between two observers, affects entanglement \cite{Rideout2012_IOP}.  The previous experiments depend on transmitting photons from a satellite.  Associated with this is the fact that there should be a relativistic correction to the mode structure of photons, such as in Laguerre-Gauss modes, or polarisation \cite{PhysRevA.101.012322}.

Satellites can also be used to probe other effects associated with entanglement.  For instance, it has been shown that the entanglement of Gaussian states \cite{Bruschi2014_PRD,Kohlrus2017_EPJ}, two mode-squeezed states \cite{Liu_2019} and multipartite W-states \cite{Wu_2020} show measurable changes in curved spacetime. Additionally, curved spacetime has  been theorised to create entanglement \cite{BALL2006550}.  An example of this is that non-uniform acceleration and changes in the gravitational field can create entanglement between modes of a single, localised quantum field, such as electromagnetic \cite{PhysRevD.85.081701} or phononic cavity modes \cite{Sab_n_2014}.  Non-uniform accelerations can also produce decoherence between spatially separated entangled systems \cite{PhysRevD.85.061701,Bruschi2014_NJP}.  Another way of testing the effect of curved spacetime on the entanglement between phononic fields is to employ Bose-Einstein condensates (BECs).  
It has been argued that entanglement between excitations of two BECs is degraded after one of them undergoes a change in the gravitational field strength~\cite{Bruschi2014_PRD}. This prediction can be tested if the two entangled BECs are initially in two separate satellites in the same orbit, after which one satellite goes into a different orbit. The effect is observable in a typical orbital manoeuvre of nanosatellites such as CanX4 and CanX5. New proposals also show that trapped BEC senors have the potential to be miniaturised \cite{Bravo2020a,Bravo2020b}. This might pave the way for other studies of fundamental physics using quantum sensors. 



Currently, experimental tests of the previously mentioned effects are limited.  
However, satellites technologies allow us to carry out experiments that will shed light on these interesting effects and allow for a deeper understanding of how gravity affects entanglement. For instance, the Space QUEST mission by ESA is designed to search for gravitational decoherence effects on an entanglement between the ground and the ISS~\cite{Joshi2018_NJP}. Similarly, the Micius satellite was used to perform an experiment on entangled photon pairs~\cite{Xu2019_S}.  These experiments showed that entanglement persists under non-inertial motion to within the resolution of the test-system. The result is in agreement with the theory which predicts that entanglement is conserved under uniform acceleration \cite{Alsing_2012}. 

\subsubsection{Tests of modified theories of gravity}
\label{subsubsec:modified_theories}

\noindent
One way of detecting modified theories of gravity is to investigate the phase picked up by photons as they travel through gravitational fields.  For instance, as discussed in the previous section, one could look at photons propagating between Earth and a satellite link~\cite{Buoninfante2020}. In particular, such experiments could detect the existence of screening scalar fields, such as Chameleon fields~\cite{khoury2004chameleon}. It is possible to determine photon trajectories for which the phase effects vanish according to the predictions of general relativity. Therefore, deviations would imply the existence of new physics.  Phase shifts of this type have been detected for massive particles, in the COW experiment as mentioned in the previous section.  Similar experiments for massless quantum particles could, in principle, be designed. 
In the context of ground-to-satellite links, the effect can be measured by preparing a superposition of two temporal modes on ground (the so called time-bin encoding) and sending these towards a satellite, where an unbalanced interferometer can measure the gravitationally induced phase shift~\cite{Rideout2012_IOP}.  

\subsubsection{High precision estimation of fundamental Physics parameters}

\noindent
Quantum resources can improve the sensitivity of detectors over classical methods. This is the field of quantum metrology, which provides a natural extension to performance improvements using quantum estimation theory. Quantum metrology has matured into a broad field with many active areas of theoretical and experimental research. Quantum states such as squeezed vacuum have been used to suppress statistical fluctuations due to shot noise, which enhances the sensitivity of detectors to faint signals~\cite{Caves1981_PRD,Pezze08}. This effect has been implemented to improve gravitational wave detectors~\cite{Aasi2013, Ligo2013_NP}. Quantum resources have also been used to precisely measure parameters that encode relativistic effects, such as proper accelerations, relative distances, time, and gravitational field strengths~\cite{PhysRevD.89.065028,ahmadi2014relativistic,2015NatSR,Fuentes2019}. Further, wave packets of light are known to evolve as they propagate near massive objects. This evolution encodes characteristic attributes of the spacetime and can be used to estimate parameters of it. Methods from relativistic quantum metrology can be readily implemented in future space experiments to estimate the spacetime parameters of the Earth including the Schwarzchild radius and Earth's equatorial angular velocity \cite{Kohlrus2019_PRA,Kohlrus2017_EPJ} with unprecedented precision. This will improve detector performances for a range of applications that include positioning, navigation, sensors, radars, and gyroscopes~\cite{Bruschi2014_PRD}. 


\subsection{Tests of foundations of Quantum Mechanics}
\label{subsec:QM_tests}

\noindent
Satellites also provide a platform for experiments that test the foundations of QM.  
One example is Wheeler's Delayed Choice Experiment, which consists of a single photon traveling into a Mach-Zehnder interferometer (MZI)~\cite{wheeler1978}. The key idea of Wheeler is the possibility of changing the interferometer configuration after the photon has already entered the MZI. The beam splitter (BS) that closes the interferometer can be removed or kept in place. Crucially, this choice is made when the photon is inside the interferomter. When the BS is present, the device shows wave-like behaviour, when the BS is removed the photons behave like a particle. Classical thinking would suggest that if the photon is either a particle or a wave, then the photon's nature should be fixed at the input of the interferometer. However, due to the ``delayed choice'', a purely classical interpretation of the process would imply a violation of causality. Several implementations of Wheeler's Gedankenexperiment have been realised on the ground~\cite{Ma2016}. Recently, the delayed-choice Gedankenexperiment was also demonstrated on space channels~\cite{vedovato2017}. In particular, the photons were prepared into a superposition of two wavepackets (for example time-bin encoding), sent towards a rapidly moving satellite in orbit, and reflected back. On the ground, the insertion or removal of the BS at the measurement apparatus was determined by a quantum random number generator and implemented after the reflection from the satellite. The experiment showed the correctness of the quantum mechanical wave-particle model also in a space domain, for propagation distances of up to 3500 km. This paves the path for further satellite communications-enabled space-based experiments to probe further fundamentals of quantum theory.

\subsection{Experiments using satellite arrays}
\noindent
A network of telescopes that work in collaboration has significantly better precision and imaging resolutions than independent telescopes. Specifically, networks with larger baselines provide a precise measurement of time differences between the arrival of a signals to each telescope. This can be used to improve image resolutions and estimate separation distances between two sources that contribute to the detected signals. This is the basis of coherent amplitude interferometry, where multiple objects in mutual close proximity can be distinguished. This feature has been demonstrated by terrestrial networks of telescopes to observe the structure of quasars~\cite{VLBI2013}, and black hole imaging~\cite{collaborat2019first}. Satellite-based networks also allow access to baselines that are significantly longer than planetary scales. This promises more precise measurements and higher resolution images of the universe. In the following sections, we review how these longer baselines can benefit multi-messenger astronomy, and search for theories beyond the standard model (BSM) of Physics and dark matter.

\subsubsection{Multi-messenger astronomy and gravitational waves}

\noindent
With the advent of the gravitational wave era \cite{AbbottPRL2016}, multi-messenger astronomy has emerged as a promising tool to measure different signals from an astronomical event~\cite{Meszaros:2019nature}. For a blackhole merger, this includes measurements of gamma rays in addition to gravitational waves. In addition, BSM theories predict the presence of low mass particles from exotic light fields (ELFs)~\cite{Curtin:2019PLB}. For example, exotic bosonic fields can give rise to axions and solitons~\cite{Helfer2019PRD}. Detection of these particles has been associated with dark matter and explanations for the hierarchy problem and the strong charge-parity (CP)   problem~\cite{Sakharov1992_SPU}. Particles from ELFs have also been proposed to be found in the vicinity of blackholes causing an effect called super-luminescence. Although high-energy events are not necessary for the production of ELFs, a significant number of ELFs must be produced for terrestrial detectors to observe them. An array of small satellites in LEO, each equipped with quantum sensors for precise measurement of signal arrival times, provides a suitably long baseline for multi-messenger astronomy of ELFs~\cite{dailey2020quantum}. 
It has also been shown that a BEC can be used as a quantum Weber bar that can detect gravitational wave signals at high frequencies~\cite{Sab_n_2014}, reaching frequencies that are not accessible by other detectors such as LIGO. A global network of BEC gravitational wave detectors could be used to improve localisation and resolution capabilities.

\subsubsection{Search for dark matter}
\label{subsubsec:AGW}
\noindent
Dark matter constitutes more than 80\% of the known universe but cannot be directly observed. Proposals for indirect measurement of dark matter search for weakly interacting massive particles, topological defects in the universe~\cite{Kawasaki:2014PRD}, or other exotic states of matter. For these proposals, the effects of dark matter are measured through the use of multi-messenger astronomy that is tuned to detect minute disparate signals~\cite{Djuna2019, Bhoonah:2020oov}. For example, detecting topological defects involves measuring variations of fundamental constants~\cite{Roberts:2017naturecomms, Derevianko:2013naturephys, Derevianko2013PRA}. These variations emerge as shifts in the atomic energy levels of particles that can be measured by monitoring their atomic frequencies. To maximise the sensitivity of detecting dark matter candidates, sensors with longer baselines are required. A network of synchronised small satellites or CubeSats would provide much greater coverage than terrestrial-based arrays. Any topological defect passing through atomic clocks within the network would lead an desynchronisation of the clocks with respect to a reference clock on the ground~\cite{Derevianko2018PRD}. BECs can also be used to search for ultralight dark-matter particles~\cite{howl2021}, constraining dark energy models~\cite{Hartley2019}, and testing physical regimes where quantum mechanics and general relativity may interact.

\section{Conclusion}
\label{sec:conclusions}

\noindent
Space enables quantum communication applications across global scales by overcoming the range limitation of current ground based networks. Creating space-based entanglement distribution links has unique theoretical and engineering challenges. Ultimately, the vision is to integrate space-based systems with current terrestrial optical networks to realise a truly global quantum internet. Significant recent progress in feasibility studies and theoretical work have helped understand and model limiting factors to guide future field demonstrations of key milestones towards this vision.

The original ideas for satellite quantum communications were developed from the 1990s with theoretical analysis and mission proposals, culminating in the landmark in-orbit demonstration by the Micius satellite. Spurred by its success, international interest in space-based quantum communications was reignited. We capture recent advances by reviewing concerted academic, governmental, and commercial efforts. 

The quantum internet is set to deliver a profound impact across the spectrum of quantum technologies. Applications of quantum communications have matured into a broad field with many active areas of theoretical and experimental research. In this review, we provide a summary of this progress and establish a roadmap to the development of a space segment of a global quantum internet. We identify key challenges, potential solutions and capabilities to develop. The principle challenge is the establishment of ground and satellite links that allow for efficient distribution and routing of quantum entanglement. We summarise readiness of different technologies for space quantum communications and highlight key milestones achieved, and missions in development, succinctly summarised in a detailed timeline.  

Recent developments in space systems, especially the rapid adoption of small satellites and CubeSats in particular, have been enthusiastically adopted by space quantum communications missions. Small satellites allow for rapid and less-costly space systems developments, which is especially important in a rapidly moving field. The quickly expanding capabilities of small satellites has been driven by miniaturisation of components, leading them to be viable platforms beyond simply educational tools. This parallel trend exists in striving to miniaturise and make more robust quantum components and devises leading to a surge of in-orbit demonstrations with CubeSats. Further developments offer the possibility of large constellation of CubeSats, providing complementary services to a smaller number of larger satellites in terms of coverage.

We provide an overview of key remaining challenges that fall within two broad areas: first, increasing the robustness of quantum signals to mitigate noise in free-space channels; second, engineering each component for space preparedness. We highlight the recent progress that aims to improve space quantum technologies by mitigating these effects. We also provide a perspective on future system-level changes that could deliver performance improvements. We conclude with how space quantum communication could support advancements across fundamental Physics.



\section{Acknowledgments}
\label{sec:acknowledgements}

\noindent
We acknowledge UKRI EPSRC Quantum Technology Hub in Quantum Communications Grant No. EP/M013472/1, EP/T001011/1, and Partnership Resource, UK Space Agency grants NSTP3-FT-063 \& NSTP3-FT2-065 (QSTP), COST Action QTSpace (CA15220), Innovate UK (AQKD \#45364, ViSatQT \#43037), and European Union grant CiViQ (No. 820466). MG \& MK acknowledge the German Space Agency DLR and funds provided by the Federal Ministry of Economics and Technology (BMWi) (grants 50WM1958 (OPTIMO) \& 50WM2055 (OPTIMO-2)). MG acknowledges the European Union Horizon 2020 R\&I programme Marie Sk{\l}odowska-Curie grant No. 894590. SM acknowledges the IAA fellowship at UoS and Craft Prospect supporting EPSRC funded research. JS acknowledges funding from SUPA. DKLO is supported by EPSRC grant EP/T517288/1. GV and PV acknowledge funding from Ministero dell'Istruzione,  dell'Universit\'{a}  e  della  Ricerca  under  initiative ``Departments of Excellence'' (Law 232/2016). MM and LM performed work at the Jet Propulsion Laboratory, California Institute of Technology under a contract with the National Aeronautics and Space Administration’s Biological and Physical Science Division.



\bibliographystyle{iet}

\end{document}